\documentclass[showpacs,prd]{revtex4}
\usepackage{amssymb}
\usepackage{amsmath}
\usepackage{graphicx}
\usepackage{bm}
\usepackage{amsfonts}

\begin{document}

\title{Derivation of transient relativistic fluid dynamics from the
Boltzmann equation}
\author{G.\ S.\ Denicol${}^{a}$, H.\ Niemi${}^{b,c}$, E.\ Moln\'{a}r$%
{}^{b,d} $, and D.\ H.\ Rischke${}^{a,b}$}
\affiliation{$^{a}$Institut f\"ur Theoretische Physik, Johann Wolfgang
Goethe-Universit\"at, Max-von-Laue-Str.\ 1, D-60438 Frankfurt am Main,
Germany}
\affiliation{$^{b}$Frankfurt Institute for Advanced Studies, Ruth-Moufang-Str.\ 1,
D-60438 Frankfurt am Main, Germany}
\affiliation{$^{c}$Department of Physics, P.O. Box 35 (YFL) FI-40014 University of
Jyv\"askyl\"a, Finland}
\affiliation{$^{d}$MTA Wigner Research Centre for Physics, H-1525 Budapest, P.O.Box 49,
Hungary}

\begin{abstract}
In this work we present a general derivation of relativistic fluid dynamics
from the Boltzmann equation using the method of moments. The main difference
between our approach and the traditional 14-moment approximation is that we
will not close the fluid-dynamical equations of motion by truncating the
expansion of the distribution function. Instead, we keep all terms in the
moment expansion. The reduction of the degrees of freedom is done by
identifying the microscopic time scales of the Boltzmann equation and
considering only the slowest ones. In addition, the equations of motion for
the dissipative quantities are truncated according to a systematic
power-counting scheme in Knudsen and inverse Reynolds number. We conclude
that the equations of motion can be closed in terms of only 14 dynamical
variables, as long as we only keep terms of second order in Knudsen and/or
inverse Reynolds number. We show that, even though the equations of motion
are closed in terms of these 14 fields, the transport coefficients carry
information about all the moments of the distribution function. In this way,
we can show that the particle-diffusion and shear-viscosity coefficients
agree with the values given by the Chapman-Enskog expansion.
\end{abstract}

\date{\today }
\maketitle


\section{Introduction}

Relativistic fluid dynamics is an effective theory to describe the
long-distance, long-time dynamics of macroscopic systems, with important
applications in relativistic heavy-ion collisions and astrophysics \cite%
{Pasi}. Relativistic fluid dynamics describes the conservation of (net)
particle number and energy-momentum, 
\begin{equation}
\partial_{\mu}N^{\mu}=0\;,\;\;\;\;\partial_{\mu}T^{\mu\nu}=0\;.
\label{delNdelT}
\end{equation}
In general, these five equations contain 14 unknown fields, the four
components of the particle 4-current $N^{\mu}$ and the ten components of the
(symmetric) energy-momentum tensor $T^{\mu\nu}$. Thus, these equations are
not closed and one needs to specify nine additional equations of motion to
solve them. The coefficients in the equations of motion (equation of state,
transport coefficients, etc.) must be determined by matching fluid dynamics
to the underlying microscopic theory. In the case of dilute gases, this is
the Boltzmann equation.

There are two widespread methods to provide additional equations of motion
from the Boltzmann equation: the Chapman-Enskog expansion and the method of
moments. In the Chapman-Enskog expansion \cite{Chapman}, the corrections to
the single-particle distribution function in local equilibrium are assumed
to be functions of the five traditional fluid-dynamical variables,
temperature, chemical potential, and the three components of the
fluid-velocity field, as well as gradients thereof. The corrections are
systematically arranged in terms of an expansion in powers of the Knudsen
number, given by the ratio of the mean-free path of the particles and a
characteristic macroscopic length scale. As is well-known, the first-order
truncation of the expansion leads to Navier-Stokes theory. Keeping second
and higher-order terms one obtains the Burnett and super-Burnett equations,
respectively \cite{Burnett}. However, it has been shown that the Burnett
equations suffer from the so-called Bobylev instability \cite{Bobylev}. In
the relativistic case, even the first-order equations, i.e., the
relativistic generalization of the Navier-Stokes equations, are unstable 
\cite{his}.

Therefore, the relativistic extension of Chapman-Enskog theory should not be
applied to derive the equations of relativistic fluid dynamics from kinetic
theory. On the other hand, the method of moments \cite{DeGroot} avoids the
above mentioned problems. The method of moments was first developed by Grad 
\cite{Grad} for non-relativistic systems. In Grad's original work, the
single-particle distribution function is expanded around its local
equilibrium value in terms of a complete set of Hermite polynomials \cite%
{Grad_Hermite}. This expansion is truncated and the distribution function is
finally expressed in terms of 13 fluid-dynamical variables: the velocity
field, the temperature, the chemical potential, the heat-conduction current,
and the shear-stress tensor. In this case the heat-conduction current and
shear-stress tensor become independent dynamical variables which satisfy
partial differential equations that describe their relaxation towards their
respective Navier-Stokes values. Grad's method is usually considered to be
independent of the Chapman-Enskog expansion. However, we emphasize that
Burnett-type equations can be obtained as the solution of Grad's equations
in the long-time limit \cite{Kremer,paperpoles}.

Nevertheless, Grad's method has one major drawback: unlike the
Chapman-Enskog expansion it lacks a small parameter, such as the Knudsen
number, in which one can do power-counting and thus systematically improve
the approximation \cite{Struchtrup_Review}. This deficiency, together with
the bad performance of Grad's method in comparison to microscopic
calculations \cite{torrilhon_Ed}, have led researchers to abandon this
approach for some time. However, recently a lot of effort has been made to
reformulate the method of moments into a more reliable tool to describe
nonequilibrium phenomena for large Knudsen numbers \cite{torrilhon_Ed}. For
instance, in Ref.\ \cite{torrilhon_R13} Grad's equations were regularized to
have a wider domain of validity in Knudsen number and then shown to be in
good agreement with microscopic calculations. Such approaches, however, were
only formulated for non-relativistic systems.

The generalization of Grad's method of moments to relativistic systems has
been pursued by several authors \cite{Chernikov}. The most widely employed
approach is due to Israel and Stewart \cite{IS}. Here, the distribution
function is expanded around its local equilibrium value in terms of a series
of (reducible) Lorentz tensors formed of particle 4-momentum $k^{\mu }$,
i.e., $1,\,k^{\mu },\,k^{\mu }k^{\nu },\,\ldots $. In Israel and Stewart's
14-moment approximation one truncates the expansion at second order in
momentum, i.e., one only keeps the tensors $1,\,k^{\mu }$, and $k^{\mu
}k^{\nu }$, with 14 unknown coefficients (the trace of $k^{\mu }k^{\nu }$ is
equal to $m^{2}$, the rest mass of the particles) to describe the
distribution function. The coefficients of the \textit{truncated} expansion
can then be uniquely related to the 14 components of the particle 4-current, 
$N^{\mu }$, and the energy-momentum tensor, $T^{\mu \nu }$, the so-called 
\textit{matching procedure}. While particle and energy-momentum conservation
(\ref{delNdelT}) are obtained from the zeroth and the first moment of the
Boltzmann equation, the additional nine equations of motion follow from the
second moment of the Boltzmann equation. However, Israel and Stewart's
theory shares the same problems of Grad's original approach: it lacks a
parameter in which one can do systematic power-counting of corrections to
the local equilibrium distribution function.

It was recently confirmed that, at least for some special problems, the
Israel-Stewart equations \cite{IS} are not in good agreement with the
numerical solution of the Boltzmann equation \cite{el,BAMPS}. Initial
attempts to improve Israel and Stewart's theory were already made in Refs.\ 
\cite{Muronga:2006zx,dkr,Betz:2008me}, but Israel and Stewart's 14-moment
approximation was still used. In this paper we demonstrate that
Israel-Stewart theory, as well as all previous attempts to improve it are
actually incomplete. The reason is that the 14-moment approximation neglects
infinitely many terms of first order in the Knudsen number. In our approach
all terms of the moment expansion are included and the exact equations of
motion for these moments are derived. These exact equations still contain
the degrees of freedom and microscopic time scales of the Boltzmann
equation. We prove that, in order to derive a causal dynamical equation for
a given dissipative current, it is necessary to resolve at least the slowest
corresponding microscopic time scale arising from the Boltzmann equation, in
agreement with the results of Ref.\ \cite{paperpoles}. Unlike in
Israel-Stewart theory, the truncation of the resulting equations of motion
in terms of only 14 dynamical variables is then implemented by a systematic
power-counting scheme in Knudsen number, $\mathrm{Kn}$, and in the ratios, $%
\mathrm{R}_{\Pi }^{-1}\equiv \left\vert \Pi \right\vert /P_{0},$ $\mathrm{R}%
_{n}^{-1}\equiv \left\vert n^{\mu }\right\vert /n_{0},$ $\mathrm{R}_{\pi
}^{-1}\equiv \left\vert \pi ^{\mu \nu }\right\vert /P_{0}$, where $\Pi $ is
the bulk viscous pressure, $n^{\mu }$ is the particle-diffusion current, $%
\pi ^{\mu \nu }$ is the shear-stress tensor, and $P_{0}$ and $n_{0}$ are the
pressure and the particle density in local equilibrium, respectively. The
ratio $\mathrm{R}_{\pi }^{-1}$ is related to the inverse Reynolds number in
non-relativistic situations. We shall in somewhat loose terminology refer to
all of them as \textquotedblleft inverse Reynolds numbers\textquotedblright\
in the following. The resulting fluid-dynamical equations and coefficients
are different from the ones obtained via the 14-moment approximation. We
calculate the numerical values of the coefficients for a massless classical
gas. We show that our values for the heat-conductivity and shear-viscosity
coefficient agree with the ones calculated via Chapman-Enskog theory \cite%
{DeGroot}.

This paper is organized as follows. In Sec.\ \ref{Boltz} we review how
fluid-dynamical variables are extracted from the Boltzmann equation. In
Sec.\ \ref{Mom_Meth} we demonstrate how to expand the single-particle
distribution function in terms of a \emph{complete, orthogonal\/} basis in
momentum space. In contrast to Israel and Stewart's non-orthogonal basis $%
1,\,k^{\mu },\,k^{\mu }k^{\nu },\ldots $, our approach uses \emph{%
irreducible tensors\/} in 4-momentum $k^{\mu }$, and is thus orthogonal. The
coefficients of the irreducible tensors in the expansion of the
single-particle distribution function are orthogonal polynomials in the
rest-frame energy and moments of the correction to the equilibrium
distribution function. Section \ref{EoM_Gen} derives an infinite set of
equations for these moments, which is still completely equivalent to the
Boltzmann equation. In Sec.\ \ref{Power} we introduce our power-counting
scheme in terms of Knudsen and inverse Reynolds numbers. Then, by
diagonalizing the linear part of the set of moment equations, we demonstrate
how to identify the slowest microscopic time scale of the Boltzmann equation
for each dissipative current. We shall derive dynamical equations for the
slowest modes, but approximate faster modes by their asymptotic solution for
long times. This will then lead, in Sec.\ \ref{Fluid_Dynamics copy}, to the
complete set of fluid-dynamical equations which contains \emph{all terms up
to second order in Knudsen and inverse Reynolds numbers}, i.e., $\mathcal{O}(%
\mathrm{Kn}^{2},\,\mathrm{R}_{i}^{-1}\mathrm{R}_{j}^{-1},\,\mathrm{Kn\,R}%
_{i}^{-1})$. In Sec.\ \ref{Applications} we first demonstrate the validity
of our approach by restricting the calculation to the 14-moment
approximation and recovering the results of Ref.\ \cite{dkr} for the
transport coefficients for the case of an ultrarelativistic, classical gas
with constant cross section. We then show how to successively improve the
expression for the transport coefficients by extending the number of moments
to $14+9\times n$. We explicitly study the cases $n=1,2,$ and 3. We end this
work with a discussion and conclusions in Sec.\ \ref{conclusions}. Various
appendices contain intermediate steps of our calculations. We use natural
units $\hbar =c=k_{B}=1$. The metric tensor is $g_{\mu \nu }=\mathrm{diag}%
\,(+,-,-,-)$.

\section{Fluid-dynamical variables from the Boltzmann equation}

\label{Boltz}

We start with the relativistic Boltzmann equation, 
\begin{equation}
k^{\mu}\partial_{\mu}f_{\mathbf{k}}=C\left[ f\right] ,  \label{Boltzmann_Eq}
\end{equation}
where $k^{\mu}=(k^{0},\mathbf{k})$, with $k^{0}=\sqrt{\mathbf{k}^{2}+m^{2}}$
and $m$ being the mass of the particles. For the collision term, we consider
only elastic two-to-two collisions with incoming momenta $k,k^{\prime}$, and
outgoing momenta $p,p^{\prime}$, 
\begin{equation}
C\left[ f\right] =\frac{1}{\nu}\int dK^{\prime}dPdP^{\prime}W_{\mathbf{kk}%
\prime\rightarrow\mathbf{pp}\prime}\left( f_{\mathbf{p}}f_{\mathbf{p}%
^{\prime}}\tilde{f}_{\mathbf{k}}\tilde{f}_{\mathbf{k}^{\prime}}-f_{\mathbf{k}%
}f_{\mathbf{k}^{\prime}}\tilde{f}_{\mathbf{p}}\tilde{f}_{\mathbf{p}%
^{\prime}}\right) ,  \label{Col_term}
\end{equation}
where $\nu$ is a symmetry factor ($=2$ for identical particles), $W_{\mathbf{%
kk}\prime\rightarrow \mathbf{pp}\prime}$ is the Lorentz-invariant transition
rate, and $dK\equiv\,gd^{3}\mathbf{k/}\left[ (2\pi)^{3}k^{0}\right] $ is the
Lorentz-invariant momentum-space volume, with $g$ being the number of
internal degrees of freedom. We introduced the notation $\tilde{f}_{\mathbf{k%
}}\equiv1-af_{\mathbf{k}}$, where $a=1$ ($a=-1$) for fermions (bosons) and $%
a=0$ for a classical gas.

In kinetic theory, the conserved particle current $N^{\mu }$ and the
energy-momentum tensor $T^{\mu \nu }$ are expressed as moments of the
single-particle distribution function, 
\begin{equation}
N^{\mu }=\left\langle k^{\mu }\right\rangle ,\text{ \ }T^{\mu \nu
}=\left\langle k^{\mu }k^{\nu }\right\rangle ,  \label{currents}
\end{equation}%
where we adopted the following notation: 
\begin{equation}
\left\langle \cdots \right\rangle \equiv \int dK\left( \cdots \right) f_{%
\mathbf{k}}\;.
\end{equation}%
The particle current and the energy-momentum tensor can be tensor-decomposed
with respect to the fluid 4-velocity $u^{\mu }$. To this end, we have to
specify the rest frame of the fluid. We introduce $u^{\mu }$ as a time-like,
normalized ($u_{\mu }u^{\mu }=1$) eigenvector of the energy-momentum tensor, 
\begin{equation}
T^{\mu \nu }u_{\nu }=\varepsilon u^{\mu },  \label{Landau}
\end{equation}%
where the eigenvalue $\varepsilon $ is the energy density, i.e., we work in
the Landau frame \cite{Landau}. Next, we divide the momentum of the
particles $k^{\mu }$ into two parts: one parallel and one orthogonal to $%
u^{\mu }$, 
\begin{equation}
k^{\mu }=E_{\mathbf{k}}u^{\mu }+k^{\left\langle \mu \right\rangle }\;,
\end{equation}%
where we defined the scalar $E_{\mathbf{k}}\equiv u_{\mu }k^{\mu }\equiv
u\cdot k$ and used the notation $A^{\left\langle \mu \right\rangle }=\Delta
_{\nu }^{\mu }A^{\nu }$, with $\Delta ^{\mu \nu }=g^{\mu \nu }-u^{\mu
}u^{\nu }$ being the projection operator onto the 3-space orthogonal to $%
u^{\mu }$.

Then, the tensor decomposition of $N^{\mu }$ and $T^{\mu \nu }$ reads 
\begin{equation}
N^{\mu }=nu^{\mu }+n^{\mu },\text{ \ }T^{\mu \nu }=\varepsilon \,u^{\mu
}u^{\nu }-\Delta ^{\mu \nu }\left( P_{0}+\Pi \right) +\pi ^{\mu \nu }\;,
\end{equation}%
where the particle density $n$, the particle-diffusion current $n^{\mu }$,
the energy density $\varepsilon $, the shear-stress tensor $\pi ^{\mu \nu }$%
, and the sum of thermodynamic pressure, $P_{0}$, and bulk viscous pressure, 
$\Pi $, are defined by 
\begin{equation}
n\equiv \left\langle E_{\mathbf{k}}\right\rangle \,,\;n^{\mu }\equiv
\left\langle k^{\left\langle \mu \right\rangle }\right\rangle
\,,\;\varepsilon \equiv \left\langle E_{\mathbf{k}}^{2}\right\rangle \,,%
\text{ }\pi ^{\mu \nu }\equiv \left\langle k^{\left\langle \mu \right.
}k^{\left. \nu \right\rangle }\right\rangle \,,\;P_{0}+\Pi \equiv -\frac{1}{3%
}\left\langle \Delta ^{\mu \nu }k_{\mu }k_{\nu }\right\rangle \;,
\label{def_hy_quan}
\end{equation}%
where $A^{\langle \mu \nu \rangle }\equiv \Delta _{\alpha \beta }^{\mu \nu
}A^{\alpha \beta }$ and $\Delta _{\alpha \beta }^{\mu \nu }\equiv \lbrack
\Delta _{\alpha }^{\mu }\Delta _{\beta }^{\nu }+\Delta _{\alpha }^{\nu
}\Delta _{\beta }^{\mu }-\left( 2/3\right) \Delta ^{\mu \nu }\Delta _{\alpha
\beta }]/2$ denotes a projector onto that part of a rank-2 tensor, which is
symmetric, orthogonal to $u^{\mu }$, and traceless.

Next, we introduce the local-equilibrium distribution function as $f_{0%
\mathbf{k}}=\left[ \exp \left( \beta _{0}\,E_{\mathbf{k}}-\alpha _{0}\right)
+a\right] ^{-1}$, where $\beta _{0}$ and $\alpha _{0}$ are the inverse
temperature and the ratio of the chemical potential to temperature,
respectively. The values of $\alpha _{0}$ and $\beta _{0}$ are determined by
the matching conditions, 
\begin{equation}
n\equiv n_{0}=\langle E_{\mathbf{k}}\rangle _{0},\;\;\;\varepsilon \equiv
\varepsilon _{0}=\left\langle E_{\mathbf{k}}^{2}\right\rangle _{0},
\label{matching}
\end{equation}%
where 
\begin{equation}
\langle \cdots \rangle _{0}\equiv \int dK\left( \cdots \right) f_{0\mathbf{k}%
}\;.
\end{equation}%
Then, the separation between thermodynamic pressure and bulk viscous
pressure is achieved as $P_{0}=-\left\langle \Delta ^{\mu \nu }k_{\mu
}k_{\nu }\right\rangle _{0}/3$ and $\Pi =-\left\langle \Delta ^{\mu \nu
}k_{\mu }k_{\nu }\right\rangle _{\delta }/3$, where 
\begin{equation}
\langle \cdots \rangle _{\delta }=\langle \cdots \rangle -\langle \cdots
\rangle _{0}\;.
\end{equation}

The fluid-dynamical conservation laws (\ref{delNdelT}) are equations of
motion for $n$, $\varepsilon$, and $u^{\mu}$, hence one needs nine
additional equations to determine the dissipative corrections $%
\Pi,\,n^{\mu}, $ and $\pi^{\mu\nu}$. In the following, we shall use the
method of moments to derive these equations.

\section{Expansion of the single-particle distribution function in terms of
irreducible tensors}

\label{Mom_Meth}

In this section, we expand the single-particle distribution $f_{\mathbf{k}}$
in terms of irreducible tensors. It is convenient to factorize the
local-equilibrium distribution function $f_{0\mathbf{k}}$ from $f_{\mathbf{k}%
}$, 
\begin{equation}
f_{\mathbf{k}}=f_{0\mathbf{k}}\left( 1+\tilde{f}_{0\mathbf{k}}\phi _{\mathbf{%
k}}\right) ,  \label{small_correc}
\end{equation}%
where $\phi _{\mathbf{k}}$ represents the deviation from local equilibrium
and is a function of $x^{\mu }$ and $k^{\mu }$, which is ultimately
determined by the solution of the Boltzmann equation (\ref{Boltzmann_Eq}).

The next step is to expand $\phi_{\mathbf{k}}$ in terms of a complete basis
of tensors formed of $k^{\mu}$ and $E_{\mathbf{k}}$. As mentioned in the
introduction, Israel and Stewart chose the following basis to expand $\phi_{%
\mathbf{k}}$: $1$, $k^{\mu}$, $k^{\mu}k^{\nu}$, $k^{\mu }k^{\nu}k^{\lambda}$%
, $\ldots$, and then truncated the expansion after the second-rank tensor $%
k^{\mu}k^{\nu}$, that is $\phi_{\mathbf{k}}=\epsilon _{\mathbf{k}}+\epsilon_{%
\mathbf{k}}^{\mu}k_{\mu}+\epsilon_{\mathbf{k}}^{\mu \nu}k_{\mu}k_{\nu}$,
where $\epsilon_{\mathbf{k}}$, $\epsilon_{\mathbf{k}}^{\mu}$, $\epsilon_{%
\mathbf{k}}^{\mu\nu}$ are the expansion coefficients \cite{IS}. Note that
these tensors are \textit{not irreducible} with respect to Lorentz
transformations $\Lambda_{\hspace*{0.1cm}\nu}^{\mu}$ that leave the fluid
4-velocity $u^{\mu}$ invariant, $\Lambda_{\hspace*{0.1cm}\nu}^{\mu}u^{%
\nu}=u^{\mu}$. As a consequence, they are also not orthogonal, see Chapter
VI, Sec.\ 2a of Ref.\ \cite{DeGroot}. Therefore, the expansion coefficients
cannot be straightforwardly obtained: in a non-orthogonal basis, this
requires in general the inversion of an infinite-dimensional matrix. Also,
this implies that the \textit{exact} form of the expansion coefficients
cannot be obtained once the expansion is \textit{truncated}. Therefore, the
approach of Israel and Stewart does not provide the complete expressions for
the expansion coefficients.

In order to avoid such problems, we expand $\phi _{\mathbf{k}}$ using the 
\textit{irreducible} tensors, 
\begin{equation}
1\,,\;k^{\left\langle \mu \right\rangle }\,,\;k^{\left\langle \mu \right.
}k^{\left. \nu \right\rangle }\,,\;k^{\left\langle \mu \right. }k^{\nu
}k^{\left. \lambda \right\rangle }\,,\ldots ,  \label{irrtens}
\end{equation}%
as a basis. It should be emphasized that these tensors form a \textit{%
complete and orthogonal\/} set, analogous to the spherical harmonics \cite%
{Anderson}. These irreducible tensors are defined by using the symmetrized
and, for $m>1$ traceless, projection orthogonal to $u^{\mu }$ as 
\begin{equation}
A^{\left\langle \mu _{1}\cdots \mu _{m}\right\rangle }\equiv \Delta _{\nu
_{1}\cdots \nu _{m}}^{\mu _{1}\cdots \mu _{m}}A^{\nu _{1}\cdots \nu _{m}}\;,
\label{Aproj}
\end{equation}%
where the projectors $\Delta _{\nu _{1}\cdots \nu _{m}}^{\mu _{1}\cdots \mu
_{m}}$ are defined in Ref.\ \cite{DeGroot}, see Appendix \ref%
{irreducible_tensors} for details. In order to obtain the irreducible
tensors (\ref{irrtens}), we apply the projection (\ref{Aproj}) to $A^{\nu
_{1}\cdots \nu _{m}}\equiv k^{\nu _{1}}\cdots k^{\nu _{m}}$. The tensors (%
\ref{irrtens}) satisfy an orthogonality condition, 
\begin{equation}
\int dK\,F_{\mathbf{k}}\,k^{\left\langle \mu _{1}\right. }\cdots k^{\left.
\mu _{m}\right\rangle }\,k_{\left\langle \nu _{1}\right. }\cdots k_{\left.
\nu _{n}\right\rangle }=\frac{m!\,\delta _{mn}}{\left( 2m+1\right) !!}%
\,\Delta _{\nu _{1}\cdots \nu _{m}}^{\mu _{1}\cdots \mu _{m}}\int dK\,F_{%
\mathbf{k}}\left( \Delta ^{\alpha \beta }k_{\alpha }k_{\beta }\right) ^{m},
\label{orthogonality1}
\end{equation}%
where $n,m=0,1,2,\ldots $, $F_{\mathbf{k}}$ is an arbitrary function of $E_{%
\mathbf{k}}$ and $\delta _{mn}$ denotes the Kronecker-delta. Using the basis
(\ref{irrtens}), $\phi _{\mathbf{k}}$ can be expanded as 
\begin{equation}
\phi _{\mathbf{k}}=\sum_{\ell =0}^{\infty }\lambda _{\mathbf{k}%
}^{\left\langle \mu _{1}\cdots \mu _{\ell }\right\rangle }\,k_{\left\langle
\mu _{1}\right. }\cdots k_{\left. \mu _{\ell }\right\rangle }\;.
\label{expansion1}
\end{equation}%
The index $\ell $ indicates the rank of the tensor $\lambda _{\mathbf{k}%
}^{\left\langle \mu _{1}\cdots \mu _{\ell }\right\rangle }$ and $\ell =0$
corresponds to the scalar $\lambda $. The coefficients $\lambda _{\mathbf{k}%
}^{\left\langle \mu _{1}\cdots \mu _{\ell }\right\rangle }$ are complicated
functions of $E_{\mathbf{k}}$ and are further expanded in terms of an
orthogonal basis of functions $P_{\mathbf{k}n}^{\left( \ell \right) }$, 
\begin{equation}
\lambda _{\mathbf{k}}^{\left\langle \mu _{1}\cdots \mu _{\ell }\right\rangle
}=\sum_{n=0}^{N_{\ell }}c_{n}^{\left\langle \mu _{1}\cdots \mu _{\ell
}\right\rangle }P_{\mathbf{k}n}^{\left( \ell \right) }\;,  \label{expansion2}
\end{equation}%
where $N_{\ell }$ is the number of functions $P_{\mathbf{k}n}^{\left( \ell
\right) }$ considered to describe the $\ell $-th rank tensor $\lambda _{%
\mathbf{k}}^{\left\langle \mu _{1}\cdots \mu _{\ell }\right\rangle }$. In
principle, $N_{\ell }$ should be infinite, however in practice, the
expansion (\ref{expansion2}) must be truncated and $N_{\ell }$ characterizes
the truncation order. The function $P_{\mathbf{k}n}^{\left( \ell \right) }$
are chosen to be polynomials of order $n$ in energy, $E_{\mathbf{k}}$, 
\begin{equation}
P_{\mathbf{k}n}^{\left( \ell \right) }=\sum_{r=0}^{n}a_{nr}^{(\ell )}E_{%
\mathbf{k}}^{r}\;,  \label{Poly}
\end{equation}%
which are constructed to satisfy the orthonormality condition 
\begin{equation}
\int dK\,\omega ^{\left( \ell \right) }\,P_{\mathbf{k}m}^{\left( \ell
\right) }P_{\mathbf{k}n}^{\left( \ell \right) }=\delta _{mn},
\label{conditions}
\end{equation}%
where $\omega ^{\left( \ell \right) }$ is defined as%
\begin{equation}
\omega ^{\left( \ell \right) }\equiv \frac{W^{\left( \ell \right) }}{\left(
2\ell +1\right) !!}\left( \Delta ^{\alpha \beta }k_{\alpha }k_{\beta
}\right) ^{\ell }f_{0\mathbf{k}}\tilde{f}_{0\mathbf{k}}\;.
\end{equation}%
The coefficients $a_{nr}^{(\ell )}$ and the normalization constants $%
W^{\left( \ell \right) }$ can be found via Gram-Schmidt orthogonalization
using the orthonormality condition (\ref{conditions}), see Appendix \ref%
{orthogonal polynomials} for details. We note that, in the limit of
massless, classical particles, the polynomials $P_{\mathbf{k}n}^{\left( \ell
\right) }$ correspond to the associated Laguerre polynomials.

Since the expansion (\ref{expansion1}) employs an orthogonal basis, the
expansion coefficients in Eq. (\ref{expansion2}) can be immediately
determined using Eqs. (\ref{orthogonality1}) and (\ref{conditions}). For $%
n\leq N_{\ell }$ they are given by 
\begin{equation}
c_{n}^{\left\langle \mu _{1}\cdots \mu _{\ell }\right\rangle }=\frac{%
W^{\left( \ell \right) }}{\ell !}\left\langle P_{\mathbf{k}n}^{\left( \ell
\right) }\text{ }k^{\left\langle \mu _{1}\right. }\cdots k^{\left. \mu
_{\ell }\right\rangle }\right\rangle _{\delta }\;.  \label{coeff}
\end{equation}%
For the sake of later convenience, these expansion coefficients are
re-expressed as linear combinations of irreducible moments of $\delta f_{%
\mathbf{k}}\equiv f_{\mathbf{k}}-f_{0\mathbf{k}}$,%
\begin{equation}
\rho _{n}^{\mu _{1}\cdots \mu _{\ell }}\equiv \left\langle E_{\mathbf{k}}^{n}%
\text{ }k^{\left\langle \mu _{1}\right. }\cdots k^{\left. \mu _{\ell
}\right\rangle }\right\rangle _{\delta }\;,  \label{rho}
\end{equation}%
such that 
\begin{equation}
\lambda _{\mathbf{k}}^{\left\langle \mu _{1}\cdots \mu _{\ell }\right\rangle
}=\sum_{n=0}^{N_{\ell }}\mathcal{H}_{\mathbf{k}n}^{\left( \ell \right) }\rho
_{n}^{\mu _{1}\cdots \mu _{\ell }},  \label{bla11}
\end{equation}%
where we defined the energy-dependent coefficients%
\begin{equation}
\mathcal{H}_{\mathbf{k}n}^{\left( \ell \right) }\equiv \frac{W^{\left( \ell
\right) }}{\ell !}\sum_{m=n}^{N_{\ell }}a_{mn}^{(\ell )}P_{\mathbf{k}%
m}^{\left( \ell \right) }\;.  \label{Hk}
\end{equation}%
Consequently, the distribution function itself can be expressed as a series
in the irreducible moments (\ref{rho}) of $\delta f_{\mathbf{k}}$,%
\begin{equation}
f_{\mathbf{k}}=f_{0\mathbf{k}}+f_{0\mathbf{k}}\tilde{f}_{0\mathbf{k}%
}\sum_{\ell =0}^{\infty }\sum_{n=0}^{N_{\ell }}\mathcal{H}_{\mathbf{k}%
n}^{\left( \ell \right) }\rho _{n}^{\mu _{1}\cdots \mu _{\ell
}}k_{\left\langle \mu _{1}\right. }\cdots k_{\left. \mu _{\ell
}\right\rangle }\;.  \label{fexpansion}
\end{equation}%
We remark that the matching conditions and the definition of the velocity
field imply that $\rho _{1}=\rho _{2}=\rho _{1}^{\mu }=0$.

\section{General equations of motion}

\label{EoM_Gen}

The time-evolution equations for the moments $\rho _{r}^{\mu _{1}\cdots \mu
_{\ell }}$ can be obtained directly from the Boltzmann equation by applying
the comoving derivative to the definition (\ref{rho}), together with the
symmetrized traceless projection, 
\begin{equation}
\dot{\rho}_{r}^{\left\langle \mu _{1}\cdots \mu _{\ell }\right\rangle
}=\Delta _{\nu _{1}\cdots \nu _{\ell }}^{\mu _{1}\cdots \mu _{\ell }}\frac{d%
}{d\tau }\int dKE_{\mathbf{k}}^{r}k^{\left\langle \nu _{1}\right. }\cdots
k^{\left. \nu _{\ell }\right\rangle }\delta f_{\mathbf{k}},
\label{time_deriv}
\end{equation}%
where $\dot{A}\equiv u^{\mu }\partial _{\mu }A\equiv dA/d\tau $ and $\dot{%
\rho}_{r}^{\left\langle \mu _{1}\cdots \mu _{\ell }\right\rangle }\equiv
\Delta _{\nu _{1}\cdots \nu _{\ell }}^{\mu _{1}\cdots \mu _{\ell }}$ $\dot{%
\rho}_{r}^{\nu _{1}\cdots \nu _{\ell }}$. Using the Boltzmann equation (\ref%
{Boltzmann_Eq}) in the form 
\begin{equation}
\delta \dot{f}_{\mathbf{k}}=-\dot{f}_{0\mathbf{k}}-E_{\mathbf{k}}^{-1}k_{\nu
}\nabla ^{\nu }f_{0\mathbf{k}}-E_{\mathbf{k}}^{-1}k_{\nu }\nabla ^{\nu
}\delta f_{\mathbf{k}}+E_{\mathbf{k}}^{-1}C\left[ f\right] \;,
\label{linBoltz}
\end{equation}%
where $\nabla _{\mu }=\Delta _{\mu }^{\nu }\partial _{\nu }$, and
substituting this expression into Eq.\ (\ref{time_deriv}), one can obtain
the \textit{exact} equations for the comoving derivatives of $\rho _{r}^{\mu
_{1}\cdots \mu _{l}}$.

Using the power-counting scheme developed in Sec.\ \ref{Power}, we will show
that, in order to derive the equations of motion for relativistic fluid
dynamics, it is sufficient to know the time-evolution equations for the
moments (\ref{rho}) up to rank two, i.e., for $\rho _{r}$, $\rho _{r}^{\mu }$%
, and $\rho _{r}^{\mu \nu }$. Similar equations could also be derived for
higher-rank irreducible moments, if needed. Thus, using Eqs.\ (\ref%
{time_deriv}) and (\ref{linBoltz}), we obtain 
\begin{align}
\dot{\rho}_{r}-C_{r-1}& =\alpha _{r}^{(0)}\theta -\frac{G_{2r}}{D_{20}}\Pi
\theta +\frac{G_{2r}}{D_{20}}\pi ^{\mu \nu }\sigma _{\mu \nu }+\frac{G_{3r}}{%
D_{20}}\partial _{\mu }n^{\mu }+\left( r-1\right) \rho _{r-2}^{\mu \nu
}\sigma _{\mu \nu }+r\rho _{r-1}^{\mu }\dot{u}_{\mu }-\nabla _{\mu }\rho
_{r-1}^{\mu }  \notag \\
& -\frac{1}{3}\left[ \left( r+2\right) \rho _{r}-\left( r-1\right) m^{2}\rho
_{r-2}\right] \theta ,  \label{Scalar_n} \\
\dot{\rho}_{r}^{\left\langle \mu \right\rangle }-C_{r-1}^{\left\langle \mu
\right\rangle }& =\alpha _{r}^{(1)}I^{\mu }+\rho _{r}^{\nu }\omega _{\left.
{}\right. \nu }^{\mu }+\frac{1}{3}\left[ \left( r-1\right) m^{2}\rho
_{r-2}^{\mu }-\left( r+3\right) \rho _{r}^{\mu }\right] \theta -\Delta
_{\lambda }^{\mu }\nabla _{\nu }\rho _{r-1}^{\lambda \nu }+r\rho _{r-1}^{\mu
\nu }\dot{u}_{\nu }  \notag \\
& +\frac{1}{5}\left[ \left( 2r-2\right) m^{2}\rho _{r-2}^{\nu }-\left(
2r+3\right) \rho _{r}^{\nu }\right] \sigma _{\nu }^{\mu }+\frac{1}{3}\left[
m^{2}r\rho _{r-1}-\left( r+3\right) \rho _{r+1}\right] \dot{u}^{\mu }  \notag
\\
& +\frac{\beta _{0}J_{r+2,1}}{\varepsilon _{0}+P_{0}}\left( \Pi \dot{u}^{\mu
}-\nabla ^{\mu }\Pi +\Delta _{\nu }^{\mu }\partial _{\lambda }\pi ^{\lambda
\nu }\right) -\frac{1}{3}\nabla ^{\mu }\left( m^{2}\rho _{r-1}-\rho
_{r+1}\right) +\left( r-1\right) \rho _{r-2}^{\mu \nu \lambda }\sigma
_{\lambda \nu },  \label{Vector_n} \\
\dot{\rho}_{r}^{\left\langle \mu \nu \right\rangle }-C_{r-1}^{\left\langle
\mu \nu \right\rangle }& =2\alpha _{r}^{(2)}\sigma ^{\mu \nu }-\frac{2}{7}%
\left[ \left( 2r+5\right) \rho _{r}^{\lambda \left\langle \mu \right.
}-m^{2}2\left( r-1\right) \rho _{r-2}^{\lambda \left\langle \mu \right. }%
\right] \sigma _{\lambda }^{\left. \nu \right\rangle }+2\rho _{r}^{\lambda
\left\langle \mu \right. }\omega _{\left. {}\right. \lambda }^{\left. \nu
\right\rangle }  \notag \\
& +\frac{2}{15}\left[ \left( r+4\right) \rho _{r+2}-\left( 2r+3\right)
m^{2}\rho _{r}+\left( r-1\right) m^{4}\rho _{r-2}\right] \sigma ^{\mu \nu }+%
\frac{2}{5}\nabla ^{\left\langle \mu \right. }\left( \rho _{r+1}^{\left. \nu
\right\rangle }-m^{2}\rho _{r-1}^{\left. \nu \right\rangle }\right)  \notag
\\
& -\frac{2}{5}\left[ \left( r+5\right) \rho _{r+1}^{\left\langle \mu \right.
}-rm^{2}\rho _{r-1}^{\left\langle \mu \right. }\right] \dot{u}^{\left. \nu
\right\rangle }-\frac{1}{3}\left[ \left( r+4\right) \rho _{r}^{\mu \nu
}-m^{2}\left( r-1\right) \rho _{r-2}^{\mu \nu }\right] \theta  \notag \\
& +\left( r-1\right) \rho _{r-2}^{\mu \nu \lambda \rho }\sigma _{\lambda
\rho }-\Delta _{\alpha \beta }^{\mu \nu }\nabla _{\lambda }\rho
_{r-1}^{\alpha \beta \lambda }+r\rho _{r-1}^{\mu \nu \lambda }\dot{u}%
_{\lambda }\;,  \label{Tensor_n}
\end{align}%
where we introduced the generalized irreducible collision terms 
\begin{equation}
C_{r}^{\left\langle \mu _{1}\cdots \mu _{\ell }\right\rangle }=\int dKE_{%
\mathbf{k}}^{r}k^{\left\langle \mu _{1}\right. }\cdots k^{\left. \mu _{\ell
}\right\rangle }C\left[ f\right] \;.  \label{General_Col_term}
\end{equation}%
We further defined the shear tensor $\sigma ^{\mu \nu }\equiv \nabla
^{\left\langle \mu \right. }u^{\left. \nu \right\rangle }$, the expansion
scalar $\theta \equiv \nabla _{\mu }u^{\mu }$, the vorticity tensor $\omega
^{\mu \nu }\equiv \left( \nabla ^{\mu }u^{\nu }-\nabla ^{\nu }u^{\mu
}\right) /2$ and introduced $I^{\mu }=\nabla ^{\mu }\alpha _{0}$. All
comoving derivatives of $\alpha _{0}$ and $\beta _{0}$ that appeared during
the derivation of the above equations were replaced using the \textit{exact}
equations obtained from the conservation laws of particle number, energy,
and momentum, 
\begin{align}
\dot{\alpha}_{0}& =\frac{1}{D_{20}}\left\{ -J_{30}\left( n_{0}\theta
+\partial _{\mu }n^{\mu }\right) +J_{20}\left[ \left( \varepsilon
_{0}+P_{0}+\Pi \right) \theta -\pi ^{\mu \nu }\sigma _{\mu \nu }\right]
\right\} ,  \label{bla1} \\
\dot{\beta}_{0}& =\frac{1}{D_{20}}\left\{ -J_{20}\left( n_{0}\theta
+\partial _{\mu }n^{\mu }\right) +J_{10}\left[ \left( \varepsilon
_{0}+P_{0}+\Pi \right) \theta -\pi ^{\mu \nu }\sigma _{\mu \nu }\right]
\right\} ,  \label{bla2} \\
\dot{u}^{\mu }& =\frac{1}{\varepsilon _{0}+P_{0}}\left( \nabla ^{\mu
}P_{0}-\Pi \dot{u}^{\mu }+\nabla ^{\mu }\Pi -\Delta _{\alpha }^{\mu
}\partial _{\beta }\pi ^{\alpha \beta }\right) .  \label{bla3}
\end{align}%
The coefficients $\alpha _{r}^{(0)}$, $\alpha _{r}^{(1)}$, and $\alpha
_{r}^{(2)}$ are functions of temperature and chemical potential and have the
general form, 
\begin{align}
\alpha _{r}^{(0)}& =\left( 1-r\right) I_{r1}-I_{r0}-\frac{1}{D_{20}}\left[
G_{2r}\left( \varepsilon _{0}+P_{0}\right) -G_{3r}n_{0}\right] \;,\text{ } \\
\alpha _{r}^{(1)}& =J_{r+1,1}-\frac{n_{0}}{\varepsilon _{0}+P_{0}}J_{r+2,1},
\\
\text{\ }\alpha _{r}^{(2)}& =I_{r+2,1}+\left( r-1\right) I_{r+2,2}\;,
\end{align}%
where we defined the thermodynamic functions 
\begin{align}
I_{nq}\left( \alpha _{0},\beta _{0}\right) & =\frac{1}{\left( 2q+1\right) !!}%
\left\langle E_{\mathbf{k}}^{n-2q}\left( -\Delta ^{\alpha \beta }k_{\alpha
}k_{\beta }\right) ^{q}\right\rangle _{0},\text{ }J_{nq}=\left. \frac{%
\partial I_{nq}}{\partial \alpha _{0}}\right\vert _{\beta _{0}}\text{ },
\label{Jnq} \\
G_{nm}& =J_{n0}J_{m0}-J_{n-1,0}J_{m+1,0}\text{ },\;\;%
\;D_{nq}=J_{n+1,q}J_{n-1,q}-J_{nq}^{2}\;.  \label{Gnq}
\end{align}

The dissipative quantities appearing in the conservation laws can be
(exactly) identified with the moments 
\begin{equation}
\rho _{0}=-\frac{3}{m^{2}}\,\Pi \;,\;\;\;\;\rho _{0}^{\mu }=n^{\mu
}\;,\;\;\;\;\;\rho _{0}^{\mu \nu }=\pi ^{\mu \nu }\;.  \label{rho0}
\end{equation}%
We note that the derivation of these general equations of motion is
independent of the form of the expansion of the single-particle distribution
we introduced in the previous section.

\section{Power counting and the reduction of dynamical variables}

\label{Power}

So far, we have derived a general expansion of the distribution function in
terms of the irreducible moments of $\delta f_{\mathbf{k}}$, as well as
exact equations of motion for these moments. There is an infinite number of
equations (labeled by the index $r$), and the equations for the moments up
to rank two, Eqs.\ (\ref{Scalar_n}) -- (\ref{Tensor_n}), contain moments of
rank higher than two. In general, one would have to solve this infinite set
of coupled equations in order to determine the time evolution of the system.
However, in the fluid-dynamical limit, it is expected that the macroscopic
dynamics of a given system simplifies, and therefore it can be described by
the conserved currents $N^{\mu }$ and $T^{\mu \nu }$ alone.

From the kinetic point of view, it is usually assumed that the validity of
the fluid-dynamical limit can be quantified by the Knudsen number, 
\begin{equation}
\mathrm{Kn}\equiv \frac{\ell _{\mathrm{micr}}}{L_{\mathrm{macr}}}\;,
\end{equation}%
where $\ell _{\mathrm{micr}}$ and $L_{\mathrm{macr}}$ are typical
microscopic and macroscopic length or time scales of the system,
respectively. The relevant macroscopic scales are usually estimated from the
gradients of fluid-dynamical quantities, while the microscopic scales are of
the order of the mean-free path or time between collisions. It is generally
assumed that when there is a clear separation of the microscopic and
macroscopic scales, i.e., when $\mathrm{Kn}\ll 1$, the microscopic details
can be safely integrated out and the dynamics of the system can be described
using only a few macroscopic fields.

Furthermore, we also expect fluid dynamics to be valid near local thermal
equilibrium, i.e., when $\delta f_{\mathbf{k}}\ll f_{0\mathbf{k}}$. We can
quantify the deviation from equilibrium in terms of the macroscopic
variables by defining a set of ratios of dissipative quantities to the
equilibrium pressure or density. These can be understood as generalizations
of the inverse Reynolds number and will be denoted as 
\begin{equation}
\mathrm{R}_{\Pi }^{-1}\equiv \frac{\left\vert \Pi \right\vert }{P_{0}}\;,%
\text{ }\mathrm{R}_{n}^{-1}\equiv \frac{\left\vert n^{\mu }\right\vert }{%
n_{0}}\;,\text{ }\mathrm{R}_{\pi }^{-1}\equiv \frac{\left\vert \pi ^{\mu \nu
}\right\vert }{P_{0}}\;.\text{\ }
\end{equation}%
Since the non-equilibrium moments are integrals of $\delta f_{\mathbf{k}}$
while the equilibrium pressure and particle density are integrals over the
equilibrium distribution function $f_{0\mathbf{k}}$, these ratios quantify
the deviations from equilibrium.

With this in mind, it is clear that these two measures, the Knudsen number
and the inverse Reynolds number, can be used to quantify the proximity of
the system to the fluid-dynamical limit. In general, these two measures are
independent of each other, e.g.\ a system can be initialized in such way
that the Knudsen number is large, but the inverse Reynolds number is small
or vice versa. When deriving transient fluid dynamics, one should not \emph{%
a priori} assume that $\mathrm{Kn}\sim \mathrm{R}_{i}^{-1}$: while the
Reynolds and Knudsen numbers are certainly related, their relation is in
principle dynamical and is precisely what we aim to find. Only for
asymptotically long times, the solutions of the dynamical equations yield $%
\mathrm{Kn}\sim \mathrm{R}_{i}^{-1}$, as will be discussed in more detail
below.

In the traditional 14-moment approximation introduced by Israel and Stewart 
\cite{IS}, the fluid-dynamical limit is implemented by a truncation of the
expansion of the distribution function, which corresponds neither to a
truncation in Knudsen nor in inverse Reynolds number. In this sense, the
domain of validity of the equations of motion obtained via the traditional
14-moment approximation is not clear, because it is not possible to
determine the order of the terms that were neglected. In order to obtain a
closed set of macroscopic equations with a clear domain of validity in both $%
\mathrm{Kn}$ and $\mathrm{R}_{i}^{-1}$, another truncation procedure is
necessary. The derivation of this is the main purpose of this section.

First, we re-write the collision terms $C_{r-1}^{\left\langle \mu _{1}\cdots
\mu _{\ell }\right\rangle }$ by linearizing the collision operator $C[f]$ in
the deviations from the equilibrium distribution functions. We then use the
moment expansion (\ref{fexpansion}) to obtain 
\begin{equation}
C_{r-1}^{\left\langle \mu _{1}\ldots \mu _{\ell }\right\rangle
}=-\sum_{n=0}^{N_{\ell }}\mathcal{A}_{rn}^{\left( \ell \right) }\,\rho
_{n}^{\mu _{1}\cdots \mu _{\ell }}+\left( \mbox{terms nonlinear in}%
\;\;\delta f\right) \;,  \label{great_formula}
\end{equation}%
where 
\begin{align}
\mathcal{A}_{rn}^{\left( \ell \right) }& =\frac{1}{\nu \left( 2\ell
+1\right) }\int dKdK^{\prime }dPdP^{\prime }W_{\mathbf{kk}\prime \rightarrow 
\mathbf{pp}\prime }f_{0\mathbf{k}}f_{0\mathbf{k}\prime }\tilde{f}_{0\mathbf{p%
}}\tilde{f}_{0\mathbf{p}\prime }  \notag \\
& \times E_{\mathbf{k}}^{r-1}k^{\left\langle \nu _{1}\right. }\cdots
k^{\left. \nu _{\ell }\right\rangle }\left( \mathcal{H}_{\mathbf{k}%
n}^{\left( \ell \right) }k_{\left\langle \nu _{1}\right. }\cdots k_{\left.
\nu _{\ell }\right\rangle }+\mathcal{H}_{\mathbf{k}^{\prime }n}^{\left( \ell
\right) }k_{\left\langle \nu _{1}\right. }^{\prime }\cdots k_{\left. \nu
_{\ell }\right\rangle }^{\prime }-\mathcal{H}_{\mathbf{p}n}^{\left( \ell
\right) }p_{\left\langle \nu _{1}\right. }\cdots p_{\left. \nu \ell
\right\rangle }-\mathcal{H}_{\mathbf{p}^{\prime }n}^{\left( \ell \right)
}p_{\left\langle \nu _{1}\right. }^{\prime }\cdots p_{\left. \nu _{\ell
}\right\rangle }^{\prime }\right) \;.  \label{integrals}
\end{align}%
The details of the derivation are relegated to Appendix \ref{collision}. The
coefficient $\mathcal{A}_{rn}^{\left( \ell \right) }$ is the $\left(
rn\right) $ element of an $\left( N_{\ell }+1\right) \times \left( N_{\ell
}+1\right) $ matrix $\mathcal{A}^{\left( \ell \right) }$ and contains all
the information of the underlying microscopic theory. We remark that, for $%
\ell =0$, the second and third rows and columns ($r,n=1,2$) and, for $\ell
=1 $, the second row and column ($r,n=1$) are zero, because the moments $%
\rho _{1}$, $\rho _{2}$, and $\rho _{1}^{\mu }$ vanish due to the definition
of the velocity field and the matching conditions, Eqs.\ (\ref{Landau}) and (%
\ref{matching}). Therefore, in order to invert $\mathcal{A}^{\left( \ell
\right) }$, for $\ell =0$, we have to exclude the second and third rows and
columns and, for $\ell =1$, the second row and column.

As already mentioned, fluid dynamics is expected to emerge when the
microscopic degrees of freedom are integrated out, and the system can be
described solely by the conserved currents. The exact equations of motion (%
\ref{Scalar_n}) -- (\ref{Tensor_n}) contain infinitely many degrees of
freedom, given by the irreducible moments of the distribution function, and
also infinitely many microscopic time scales, related to the coefficients $%
\mathcal{A}^{\left( \ell \right)}_{rn}$. As was argued in Ref.\ \cite%
{paperpoles}, the slowest microscopic time scale should dominate the
dynamics at long times, i.e., in the fluid-dynamical limit. In order to
extract the relevant relaxation scales, we have to determine the normal
modes of Eqs.\ (\ref{Scalar_n}) -- (\ref{Tensor_n}), i.e., we diagonalize
the part which is linear in the irreducible moments $\rho_r^{\mu_1 \cdots
\mu_\ell}$. These are the linear terms on the left-hand sides arising from
Eq.\ (\ref{great_formula}) and the first terms on the right-hand sides. The
nonlinear terms from Eq.\ (\ref{great_formula}) as well as the remaining
terms on the right-hand sides, which are nonlinear in the moments or are
gradients of moments, are not considered in the diagonalization procedure.
Identifying and separating the microscopic time scales of the Boltzmann
equation is also the basic step for obtaining general relations between the
irreducible moments and the dissipative currents and, as we shall see,
closing the equations of motion in terms of $N^{\mu }$ and $T^{\mu \nu}$.

For this purpose, we shall introduce the matrix $\Omega ^{\left( \ell
\right) }$ which diagonalizes $\mathcal{A}^{\left( \ell \right) }$, 
\begin{equation}
\left( \Omega ^{-1}\right) ^{\left( \ell \right) }\mathcal{A}^{\left( \ell
\right) }\Omega ^{\left( \ell \right) }=\mathrm{diag}\left( \chi
_{0}^{\left( \ell \right) },\ldots ,\chi _{j}^{\left( \ell \right) },\ldots
\right) ,  \label{verygood}
\end{equation}%
where $\chi _{j}^{\left( \ell \right) }$ are the eigenvalues of $\mathcal{A}%
^{\left( \ell \right) }$. We further define the tensors $X_{i}^{\mu
_{1}\cdots \mu _{\ell }}$ as 
\begin{equation}
X_{i}^{\mu _{1}\cdots \mu _{\ell }}\equiv \sum_{j=0}^{N_{\ell }}\left(
\Omega ^{-1}\right) _{ij}^{\left( \ell \right) }\rho _{j}^{\mu _{1}\cdots
\mu _{\ell }}.  \label{definitionX}
\end{equation}%
These are the eigenmodes of the linearized Boltzmann equation. Multiplying
Eq.\ (\ref{great_formula}) with $\left( \Omega ^{-1}\right) ^{(\ell )}$ from
the left and using Eqs.\ (\ref{verygood}) and (\ref{definitionX}) we obtain 
\begin{equation}
\sum_{j=0}^{N_{\ell }}\left( \Omega ^{-1}\right) _{ij}^{\left( \ell \right)
}C_{j-1}^{\left\langle \mu _{1}\cdots \mu _{\ell }\right\rangle }=-\chi
_{i}^{\left( \ell \right) }X_{i}^{\mu _{1}\cdots \mu _{\ell }}+\left( %
\mbox{terms nonlinear in}\text{ }\delta f\right) \;.  \label{Checkthisout}
\end{equation}%
where we do not sum over the index $i$ on the right-hand side of the
equation. Then we multiply Eqs.\ (\ref{Scalar_n}) -- (\ref{Tensor_n}) with $%
\left( \Omega ^{-1}\right) _{ir}^{\left( \ell \right) }$ and sum over $r$.
Using Eq.\ (\ref{Checkthisout}), we obtain the equations of motion for the
variables $X_{i}^{\mu _{1}\cdots \mu _{\ell }}$, 
\begin{align}
\dot{X}_{i}+\chi _{i}^{\left( 0\right) }X_{i}& =\beta _{i}^{(0)}\theta
+\left( \mbox{higher-order terms}\right) ,  \notag \\
\dot{X}_{i}^{\left\langle \mu \right\rangle }+\chi _{i}^{\left( 1\right)
}X_{i}^{\mu }& =\beta _{i}^{(1)}I^{\mu }+\left( 
\mbox{higher-order
    terms}\right) ,  \notag \\
\dot{X}_{i}^{\left\langle \mu \nu \right\rangle }+\chi _{i}^{\left( 2\right)
}X_{i}^{\mu \nu }& =\beta _{i}^{(2)}\sigma ^{\mu \nu }+\left( %
\mbox{higher-order terms}\right) ,  \label{help formulas}
\end{align}%
where we introduced the coefficients 
\begin{equation}
\beta _{i}^{(0)}=\sum_{j=0,\neq 1,2}^{N_{0}}\left( \Omega ^{-1}\right)
_{ij}^{\left( 0\right) }\alpha _{j}^{\left( 0\right) },\text{ \ \ }\beta
_{i}^{(1)}=\sum_{j=0,\neq 1}^{N_{1}}\left( \Omega ^{-1}\right) _{ij}^{\left(
1\right) }\alpha _{j}^{\left( 1\right) },\text{ \ \ }\beta
_{i}^{(2)}=2\sum_{j=0}^{N_{2}}\left( \Omega ^{-1}\right) _{ij}^{\left(
2\right) }\alpha _{j}^{\left( 2\right) }.
\end{equation}%
With \textquotedblleft higher-order terms\textquotedblright\ in Eqs.\ (\ref%
{help formulas}) we refer to the terms nonlinear in $\delta f$ from Eq.\ (%
\ref{Checkthisout}) as well as to the nonlinear and gradient terms on the
right-hand sides of Eqs.\ (\ref{Scalar_n}) -- (\ref{Tensor_n}). As expected,
the equations of motion for the tensors $X_{i}^{\mu _{1}\cdots \mu _{\ell }}$
decouple in the linear regime. Without loss of generality, we order the
tensors $X_{r}^{\mu _{1}\cdots \mu _{\ell }}$ according to increasing $\chi
_{r}^{\left( \ell \right) }$, e.g., in such a way that $\chi _{r}^{\left(
\ell \right) }<\chi _{r+1}^{\left( \ell \right) }$, $\forall $ $\ell $.

By diagonalizing Eqs.\ (\ref{Scalar_n}) -- (\ref{Tensor_n}) we were able to
identify the microscopic time scales of the Boltzmann equation given by the
inverse of the coefficients $\chi _{r}^{\left( \ell \right) }$. It is clear
that, if the nonlinear terms in Eqs.\ (\ref{help formulas}) are small
enough, each tensor $X_{r}^{\mu _{1}\cdots \mu _{\ell }}$ relaxes
independently to its respective asymptotic value, given by the first term on
the right-hand sides of Eqs.\ (\ref{help formulas}) (divided by the
corresponding $\chi _{r}^{\left( \ell \right) }$), on a time scale $\sim
1/\chi _{r}^{\left( \ell \right) }$. We will refer to these asymptotic
solutions as Navier-Stokes values. By neglecting all these relaxation
scales, i.e., taking the limit $\chi _{r}^{\left( \ell \right) }\rightarrow
\infty $ with $\beta _{r}^{(\ell )}/\chi _{r}^{\left( \ell \right) }$ fixed,
all irreducible moments $\rho _{r}^{\mu _{1}\cdots \mu _{\ell }}$ become
proportional to gradients of $\alpha _{0}$, $\beta _{0}$, and $u^{\mu }$,
and we obtain a Chapman-Enskog-type solution, which at first order in the
Knudsen number results in the relativistic Navier-Stokes equations of fluid
dynamics. As already mentioned in the introduction, this type of solution is
unstable and acausal, hence it cannot serve as a proper description of
relativistic fluids.

The solution for this problem was also mentioned in the introduction. To
obtain causal and stable equations one must take into account the
characteristic times within which the bulk viscous pressure, the
particle-diffusion current, and the shear-stress tensor relax towards their
asymptotic Navier-Stokes values. As shown in Ref.\ \cite{paperpoles}, in the
fluid-dynamical limit these are given by the slowest microscopic time scales
of the underlying microscopic theory, i.e., the fast relaxation scales are
not expected to contribute.

In practice, this is implemented by assuming that only the slowest modes
with rank $2$ and smaller, $X_{0}$, $X_{0}^{\mu }$, and $X_{0}^{\mu\nu}$,
remain in the transient regime and satisfy the partial differential
equations (\ref{help formulas}), 
\begin{align}
\dot{X}_{0}+\chi^{\left( 0\right) }_{0}X_{0}& =\beta^{(0)}_{0}\theta +\left( %
\mbox{higher-order terms}\right)\; ,  \notag \\
\dot{X}_{0}^{\left\langle \mu \right\rangle }+\chi^{\left( 1\right)}_{0}
X_{0}^{\mu }& =\beta_{0}^{(1)} I^{\mu }+\left( \mbox{higher-order
    terms}\right)\; ,  \notag \\
\dot{X}_{0}^{\left\langle \mu \nu \right\rangle }+\chi^{\left( 2\right)
}_{0}X_{0}^{\mu \nu }& =\beta_0^{(2)} \sigma ^{\mu \nu }+\left( %
\mbox{higher-order terms}\right)\; ,  \label{47}
\end{align}
while the modes described by faster relaxation scales, i.e., $X_{r}$, $%
X_{r}^{\mu }$, and $X_{r}^{\mu \nu }$, for any $r$ \textit{larger than} $0$,
will be approximated by their asymptotic solutions, 
\begin{align}
X_{r}& \simeq \frac{\beta_r^{(0)}}{\chi^{\left( 0\right) }_{r}}\theta +\left(%
\mbox{higher-order terms} \right)\; ,  \notag \\
X_{r}^{\mu }& \simeq \frac{\beta_r^{(1)}}{\chi^{\left( 1\right) }_{r}}I^{\mu
}+\left(\mbox{higher-order terms}\right)\; ,  \notag \\
X_{r}^{\mu \nu }& \simeq \frac{\beta_r^{(2)}}{\chi^{\left( 2\right)}_{r}}
\sigma ^{\mu \nu }+\left( \mbox{higher-order terms}\right)\; .  \label{what}
\end{align}
While this approximation is similar to the Chapman-Enskog expansion, Eqs.\ (%
\ref{47}) go beyond the Chapman-Enskog expansion by including the transient
dynamics.

Note that, for\ $r\geq 1$, $X_{r}$, $X_{r}^{\mu }$, and $X_{r}^{\mu \nu }$
are of first order in Knudsen number, $\mathcal{O}(\mathrm{Kn})$. The reason
is that the gradient terms $\theta ,\,I^{\mu },$ and $\sigma ^{\mu \nu }$
are proportional to $L_{\mathrm{macr}}^{-1}$, while $1/\chi _{r}^{(\ell )}$
is proportional to $\ell _{\mathrm{micr}}$. The coefficients $\beta
_{r}^{(\ell )}$ are simply functions of the thermodynamic variables $\alpha
_{0},\,\beta _{0}$, and thus of order $\mathcal{O}(1)$.

Furthermore, in order to obtain the traditional equations of fluid dynamics
given in terms of the conserved currents, there should not appear any tensor 
$X_{r}^{\mu \nu \lambda \ldots }$ with rank higher than $2$. Neglecting such
tensors can be justified by proving that they have asymptotic solutions
which are at least $\mathcal{O}(\mathrm{Kn}^{2},\mathrm{Kn}\mathrm{R}%
_{i}^{-1})$, i.e., beyond the order we consider here.

Equations (\ref{what}) enable us to approximate the irreducible moments that
do not appear in the conserved currents in terms of those that do occur,
namely the particle-diffusion current, the bulk viscous pressure, and the
shear-stress tensor. We now show how to do this. We first invert Eq.\ (\ref%
{definitionX}),%
\begin{equation}
\rho _{i}^{\mu _{1}\cdots \mu _{\ell }}=\sum_{j=0}^{N_{\ell }}\Omega
_{ij}^{\left( \ell \right) }X_{j}^{\mu _{1}\cdots \mu _{\ell }}\;,
\label{flamengo}
\end{equation}%
then, using Eqs.\ (\ref{what}), we obtain 
\begin{align}
\rho _{i}& \simeq \Omega _{i0}^{\left( 0\right)
}X_{0}+\sum_{j=3}^{N_{0}}\Omega _{ij}^{\left( 0\right) }\frac{\beta
_{j}^{(0)}}{\chi _{j}^{\left( 0\right) }}\,\theta =\Omega _{i0}^{\left(
0\right) }X_{0}+\mathcal{O}(\mathrm{Kn})\;,  \notag \\
\rho _{i}^{\mu }& \simeq \Omega _{i0}^{\left( 1\right) }X_{0}^{\mu
}+\sum_{j=2}^{N_{1}}\Omega _{ij}^{\left( 1\right) }\frac{\beta _{j}^{(1)}}{%
\chi _{j}^{\left( 1\right) }}\,I^{\mu }=\Omega _{i0}^{\left( 1\right)
}X_{0}^{\mu }+\mathcal{O}(\mathrm{Kn})\;,  \notag \\
\rho _{i}^{\mu \nu }& \simeq \Omega _{i0}^{\left( 2\right) }X_{0}^{\mu \nu
}+\sum_{j=1}^{N_{2}}\Omega _{ij}^{\left( 2\right) }\frac{\beta _{j}^{(2)}}{%
\chi _{j}^{\left( 2\right) }}\,\sigma ^{\mu \nu }=\Omega _{i0}^{\left(
2\right) }X_{0}^{\mu \nu }+\mathcal{O}(\mathrm{Kn})\;.  \label{flamengo2}
\end{align}%
Here, we indicated that the contribution from the modes $X_{r},X_{r}^{\mu },$
and $X_{r}^{\mu \nu }$ for $r\geq 1$ is of order $\mathcal{O}(\mathrm{Kn})$.

Taking $i=0$ in the above equations and, without loss of generality, setting 
$\Omega _{00}^{\left( \ell \right) }=1$, we obtain from Eqs.\ (\ref{rho0})
the relations 
\begin{align}
X_{0}& \simeq -\frac{3}{m^{2}}\Pi -\sum_{j=3}^{N_{0}}\Omega _{0j}^{\left(
0\right) }\frac{\beta _{j}^{(0)}}{\chi _{j}^{\left( 0\right) }}\,\theta \;, 
\notag \\
X_{0}^{\mu }& \simeq n^{\mu }-\sum_{j=2}^{N_{1}}\Omega _{0j}^{\left(
1\right) }\frac{\beta _{j}^{(1)}}{\chi _{j}^{\left( 1\right) }}\,I^{\mu }\;,
\notag \\
X_{0}^{\mu \nu }& \simeq \pi ^{\mu \nu }-\sum_{j=1}^{N_{2}}\Omega
_{0j}^{\left( 2\right) }\frac{\beta _{j}^{(2)}}{\chi _{j}^{\left( 2\right) }}%
\,\sigma ^{\mu \nu }\;.  \label{flamengo3}
\end{align}%
Substituting Eqs.\ (\ref{flamengo3}) into Eqs.\ (\ref{flamengo2}), 
\begin{align}
\frac{m^{2}}{3}\,\rho _{i}& \simeq -\Omega _{i0}^{\left( 0\right) }\Pi
-\left( \zeta _{i}-\Omega _{i0}^{\left( 0\right) }\zeta _{0}\right) \theta
=-\Omega _{i0}^{\left( 0\right) }\Pi +\mathcal{O}(\mathrm{Kn}),  \notag \\
\rho _{i}^{\mu }& \simeq \Omega _{i0}^{\left( 1\right) }n^{\mu }+\left(
\kappa _{n\,i}-\Omega _{i0}^{\left( 1\right) }\kappa _{n\,0}\right) I^{\mu
}=\Omega _{i0}^{\left( 1\right) }n^{\mu }+\mathcal{O}(\mathrm{Kn})\;,  \notag
\\
\rho _{i}^{\mu \nu }& \simeq \Omega _{i0}^{\left( 2\right) }\pi ^{\mu \nu
}+2\left( \eta _{i}-\Omega _{i0}^{\left( 2\right) }\eta _{0}\right) \sigma
^{\mu \nu }=\Omega _{i0}^{\left( 2\right) }\pi ^{\mu \nu }+\mathcal{O}(%
\mathrm{Kn})\;,  \notag \\
\rho _{i}^{\mu \nu \lambda \cdots }& \simeq \mathcal{O}(\mathrm{Kn}^{2},%
\mathrm{Kn}\,\mathrm{R}_{i}^{-1})\;.  \label{OMG}
\end{align}%
To obtain Eqs.\ (\ref{OMG}), we further used that $X_{r}^{\mu _{1}\cdots \mu
_{\ell }}\sim \mathcal{O}(\mathrm{Kn}^{2},\mathrm{Kn}\,\mathrm{R}_{i}^{-1})$
for $\ell \geq 3$, and defined the transport coefficients 
\begin{equation}
\zeta _{i}=\frac{m^{2}}{3}\sum_{r=0,\neq 1,2}^{N_{0}}\tau _{ir}^{\left(
0\right) }\alpha _{r}^{\left( 0\right) }\;,\ \ \kappa _{n\,i}=\sum_{r=0,\neq
1}^{N_{1}}\tau _{ir}^{\left( 1\right) }\alpha _{r}^{\left( 1\right) }\;,\ \
\eta _{i}=\sum_{r=0}^{N_{2}}\tau _{ir}^{\left( 2\right) }\alpha _{r}^{\left(
2\right) }\;,  \label{trans_coeffs}
\end{equation}%
where we introduced the inverse of $\mathcal{A}^{\left( \ell \right) }$, $%
\tau ^{\left( \ell \right) }\equiv \left( \mathcal{A}^{-1}\right) ^{\left(
\ell \right) }$ and used the relation, 
\begin{equation*}
\tau _{in}^{\left( \ell \right) }=\sum_{m=0}^{N_{\ell }}\Omega _{im}^{\left(
\ell \right) }\frac{1}{\chi _{m}^{\left( \ell \right) }}\left( \Omega
^{-1}\right) _{mn}^{\left( \ell \right) }\;.
\end{equation*}%
In the next subsection, we shall identify the coefficients $\zeta _{0}$, $%
\kappa _{n\,0}$, and $\eta _{0}$ as the bulk-viscosity, particle-diffusion,
and shear-viscosity coefficients, respectively.

So far we have proved that, by taking into account only the slowest
relaxation time scales, \textit{all} irreducible moments $\rho _{i}^{\mu \nu
\lambda \cdots }$ of the deviation of the single-particle distribution
function from the equilibrium one can be related, \textit{up to first order
in Knudsen number}, $\mathcal{O}(\mathrm{Kn})$, to the dissipative currents, 
$\Pi $, $n^{\mu }$, and $\pi ^{\mu \nu }$. This demonstrates that in this
limit, it is possible to reduce the number of dynamical variables in Eqs.\ (%
\ref{Scalar_n}) -- (\ref{Tensor_n}) to quantities appearing in the conserved
currents. This will be explicitly shown in the next section.

We remark that similar relations between the irreducible moments and the
dissipative currents can also be obtained with the 14-moment approximation,
but with a different set of proportionality coefficients. However, in the
traditional 14-moment approximation such relations are obtained by
explicitly truncating the moment expansion (\ref{fexpansion}) and, as a
result, they are not of a definite order in powers of Knudsen number. This
is the reason why the 14-moment approximation does not give rise to
equations of motion with a definite domain of validity in Knudsen and
inverse Reynolds numbers.

Note, however, that the relations (\ref{OMG}) are only valid for the moments 
$\rho _{r}^{\mu \nu \lambda \cdots }$ with positive $r$. This is not a
problem since similar relations can also be obtained for the irreducible
moments with negative $r$. We expect the expansion\ (\ref{fexpansion}) to be
complete and, therefore, any moment that does not appear in this expansion
must be linearly related to those that do appear. This means that, using the
moment expansion, Eq.\ (\ref{fexpansion}), it is possible to express the
moments with negative $r$ in terms of the ones with positive $r$.
Substituting Eq.\ (\ref{fexpansion}) into Eq. (\ref{rho}) and using Eq.\ (%
\ref{orthogonality1}), we obtain 
\begin{equation}
\rho _{-r}^{\nu _{1}\cdots \nu _{\ell }}=\sum_{n=0}^{N_{\ell }}\mathcal{F}%
_{rn}^{\left( \ell \right) }\rho _{n}^{\nu _{1}\cdots \nu _{\ell }},
\end{equation}%
where we defined the following thermodynamic integral%
\begin{equation}
\mathcal{F}_{rn}^{\left( \ell \right) }=\frac{\ell !}{\left( 2\ell +1\right)
!!}\int dK\text{ }f_{0\mathbf{k}}\tilde{f}_{0\mathbf{k}}E_{\mathbf{k}}^{-r}%
\mathcal{H}_{\mathbf{k}n}^{\left( \ell \right) }\left( \Delta ^{\alpha \beta
}k_{\alpha }k_{\beta }\right) ^{\ell }.
\end{equation}%
Therefore, Eqs.\ (\ref{OMG}) lead to%
\begin{align}
\rho _{-r}& =-\frac{3}{m^{2}}\,\gamma _{r}^{(0)}\Pi +\mathcal{O}(\mathrm{Kn}%
)\;,\text{ \ }  \notag \\
\rho _{-r}^{\mu }& =\gamma _{r}^{(1)}n^{\mu }+\mathcal{O}(\mathrm{Kn})\;,%
\text{ \ }  \notag \\
\rho _{-r}^{\mu \nu }& =\gamma _{r}^{(2)}\pi ^{\mu \nu }+\mathcal{O}(\mathrm{%
Kn})\;,\text{ \ }  \notag \\
\rho _{-r}^{\mu \nu \cdots }& =\mathcal{O}(\mathrm{Kn}^{3})\;,  \label{Epa}
\end{align}%
where we introduced the coefficients%
\begin{equation}
\gamma _{r}^{(0)}=\sum_{n=0,\neq 1,2}^{N_{0}}\mathcal{F}_{rn}^{\left(
0\right) }\Omega _{n0}^{\left( 0\right) }\;,\;\;\;\gamma
_{r}^{(1)}=\sum_{n=0,\neq 1}^{N_{1}}\mathcal{F}_{rn}^{\left( 1\right)
}\Omega _{n0}^{\left( 1\right) }\;,\;\;\;\gamma _{r}^{(2)}=\sum_{n=0}^{N_{2}}%
\mathcal{F}_{rn}^{\left( 2\right) }\Omega _{n0}^{\left( 2\right) }\;.
\end{equation}

\section{Complete fluid-dynamical equations to second order}

\label{Fluid_Dynamics copy}

Now we are ready to close Eqs.\ (\ref{Scalar_n}) -- (\ref{Tensor_n}) in
terms of the dissipative currents appearing in $N^{\mu }$ and $T^{\mu \nu }$
and derive the fluid-dynamical equations of motion. For this purpose, it is
convenient to use the inverse of $\mathcal{A}^{\left( \ell \right) }$, $\tau
^{\left( \ell \right) }=\left( \mathcal{A}^{-1}\right) ^{\left( \ell \right)
}$, which naturally satisfies $\tau ^{\left( \ell \right) }\mathcal{A}%
^{\left( \ell \right) }=\openone$. Hence, it is straightforward to rewrite
Eq.\ (\ref{great_formula}) as%
\begin{equation}
\sum_{j=0}^{N_{\ell }}\tau _{ij}^{\left( \ell \right) }C_{j-1}^{\left\langle
\mu _{1}\cdots \mu _{\ell }\right\rangle }=-\rho _{i}^{\mu _{1}\cdots \mu
_{\ell }}+\left( \mbox{terms nonlinear in}\;\delta f\right) \;.  \label{Yo}
\end{equation}%
Then we multiply Eqs.\ (\ref{Scalar_n}), (\ref{Vector_n}), and (\ref%
{Tensor_n}) by $\tau _{nr}^{\left( \ell \right) }$, sum over $r$, and
substitute Eq.\ (\ref{Yo}). Next, we use Eqs.\ (\ref{OMG}) and (\ref{Epa})
to replace all irreducible moments $\rho _{i}^{\mu _{1}\cdots \mu _{\ell }}$
appearing in the equations by the fluid-dynamical variables. Additionally,
all covariant time derivatives of $\alpha _{0}$, $\beta _{0}$, and $u^{\mu }$
are replaced by spatial gradients of fluid-dynamical variables using the
conservation laws in the form shown in Eqs.\ (\ref{bla1}), (\ref{bla2}), and
(\ref{bla3}). The resulting equations of motion are formally given as%
\begin{align}
\tau _{\Pi }\dot{\Pi}+\Pi & =-\zeta \theta +\mathcal{J}+\mathcal{K}+\mathcal{%
R}\;,  \notag \\
\tau _{n}\dot{n}^{\left\langle \mu \right\rangle }+n^{\mu }& =\kappa
_{n}I^{\mu }+\mathcal{J}^{\mu }+\mathcal{K}^{\mu }+\mathcal{R}^{\mu }\text{ }%
,  \notag \\
\tau _{\pi }\dot{\pi}^{\left\langle \mu \nu \right\rangle }+\pi ^{\mu \nu }&
=2\eta \sigma ^{\mu \nu }+\mathcal{J}^{\mu \nu }+\mathcal{K}^{\mu \nu }+%
\mathcal{R}^{\mu \nu }\;.  \label{Final}
\end{align}%
We remark that in order to derive these equations of motion, it is necessary
to use Eq.\ (\ref{verygood}) in the following form, 
\begin{equation}
\sum_{j=0}^{N_{\ell }}\mathcal{\tau }_{ij}^{\left( \ell \right) }\Omega
_{jm}^{\left( \ell \right) }=\Omega _{im}^{\left( \ell \right) }\,\frac{1}{%
\chi _{m}^{\left( \ell \right) }}\;.
\end{equation}%
In the above equations of motion all nonlinear terms and couplings to other
currents were collected in the tensors $\mathcal{J}$, $\mathcal{K}$, $%
\mathcal{R}$, $\mathcal{J}^{\mu }$, $\mathcal{K}^{\mu }$, $\mathcal{R}^{\mu
} $, $\mathcal{J}^{\mu \nu }$, $\mathcal{K}^{\mu \nu }$, and $\mathcal{R}%
^{\mu \nu }$. The tensors $\mathcal{J}$, $\mathcal{J}^{\mu }$, and $\mathcal{%
J}^{\mu \nu }$ contain all terms of first order in Knudsen and inverse
Reynolds numbers,%
\begin{align}
\mathcal{J}& =-\ell _{\Pi n}\nabla \cdot n-\tau _{\Pi n}n\cdot F-\delta
_{\Pi \Pi }\Pi \theta -\lambda _{\Pi n}n\cdot I+\lambda _{\Pi \pi }\pi ^{\mu
\nu }\sigma _{\mu \nu }\;,  \notag \\
\mathcal{J}^{\mu }& =-n_{\nu }\omega ^{\nu \mu }-\delta _{nn}n^{\mu }\theta
-\ell _{n\Pi }\nabla ^{\mu }\Pi +\ell _{n\pi }\Delta ^{\mu \nu }\nabla
_{\lambda }\pi _{\nu }^{\lambda }+\tau _{n\Pi }\Pi F^{\mu }-\tau _{n\pi }\pi
^{\mu \nu }F_{\nu }  \notag \\
& -\lambda _{nn}n_{\nu }\sigma ^{\mu \nu }+\lambda _{n\Pi }\Pi I^{\mu
}-\lambda _{n\pi }\pi ^{\mu \nu }I_{\nu }\;,  \notag \\
\mathcal{J}^{\mu \nu }& =2\pi _{\lambda }^{\left\langle \mu \right. }\omega
^{\left. \nu \right\rangle \lambda }-\delta _{\pi \pi }\pi ^{\mu \nu }\theta
-\tau _{\pi \pi }\pi ^{\lambda \left\langle \mu \right. }\sigma _{\lambda
}^{\left. \nu \right\rangle }+\lambda _{\pi \Pi }\Pi \sigma ^{\mu \nu
}\;-\tau _{\pi n}n^{\left\langle \mu \right. }F^{\left. \nu \right\rangle } 
\notag \\
& +\ell _{\pi n}\nabla ^{\left\langle \mu \right. }n^{\left. \nu
\right\rangle }+\lambda _{\pi n}n^{\left\langle \mu \right. }I^{\left. \nu
\right\rangle }\;.  \label{14_moment_terms}
\end{align}%
where we defined $F^{\mu }=\nabla ^{\mu }P_{0}$. In principle, one could
replace this quantity by the acceleration $\dot{u}^{\mu }$ using Eq.\ (\ref%
{bla3}). The tensors $\mathcal{K}$, $\mathcal{K}^{\mu }$, and $\mathcal{K}%
^{\mu \nu }$ contain all terms of second order in Knudsen number,%
\begin{align}
\mathcal{K}& =\zeta _{1}\,\omega _{\mu \nu }\omega ^{\mu \nu }+\zeta
_{2}\,\sigma _{\mu \nu }\sigma ^{\mu \nu }+\zeta _{3}\,\theta ^{2}+\zeta
_{4}\,I\cdot I+\zeta _{5}\,F\cdot F+\zeta _{6}\,I\cdot F+\zeta _{7}\,\nabla
\cdot I+\zeta _{8}\,\nabla \cdot F,  \notag \\
\mathcal{K}^{\mu }& =\kappa _{1}\sigma ^{\mu \nu }I_{\nu }+\kappa _{2}\sigma
^{\mu \nu }F_{\nu }+\kappa _{3}I^{\mu }\theta +\kappa _{4}F^{\mu }\theta
+\kappa _{5}\omega ^{\mu \nu }I_{\nu }+\kappa _{6}\Delta _{\lambda }^{\mu
}\partial _{\nu }\sigma ^{\lambda \nu }+\kappa _{7}\nabla ^{\mu }\theta , 
\notag \\
\mathcal{K}^{\mu \nu }& =\eta _{1}\omega _{\lambda }^{\left. {}\right.
\left\langle \mu \right. }\omega ^{\left. \nu \right\rangle \lambda }+\eta
_{2}\theta \sigma ^{\mu \nu }+\eta _{3}\sigma ^{\lambda \left\langle \mu
\right. }\sigma _{\lambda }^{\left. \nu \right\rangle }+\eta _{4}\sigma
_{\lambda }^{\left\langle \mu \right. }\omega ^{\left. \nu \right\rangle
\lambda }  \notag \\
& +\eta _{5}I^{\left\langle \mu \right. }I^{\left. \nu \right\rangle }+\eta
_{6}F^{\left\langle \mu \right. }F^{\left. \nu \right\rangle }+\eta
_{7}I^{\left\langle \mu \right. }F^{\left. \nu \right\rangle }+\eta
_{8}\nabla ^{\left\langle \mu \right. }I^{\left. \nu \right\rangle }+\eta
_{9}\nabla ^{\left\langle \mu \right. }F^{\left. \nu \right\rangle }.
\end{align}%
It is important to remark that among the terms of $\mathcal{O}(\mathrm{Kn}%
^{2})$ is a term $\omega _{\lambda }^{\text{ \ }\left\langle \mu \right.
}\omega ^{\left. \nu \right\rangle \lambda }$. Such a term was believed not
to exist in a derivation of fluid dynamics from the Boltzmann equation and
was therefore speculated to be of quantum nature \cite{BRSSS}. From our
derivation of fluid dynamics, one can see that this is not the case: it
simply emerges from a proper truncation of the single-particle distribution
function. The tensors $\mathcal{R}$, $\mathcal{R}^{\mu }$, and $\mathcal{R}%
^{\mu \nu }$ contain all terms of second order in inverse Reynolds number,%
\begin{align}
\mathcal{R}& =\text{ }\varphi _{1}\Pi ^{2}+\varphi _{2}n\cdot n+\varphi
_{3}\pi _{\mu \nu }\pi ^{\mu \nu },  \notag \\
\mathcal{R}^{\mu }& =\varphi _{4}n_{\nu }\pi ^{\mu \nu }+\varphi _{5}\Pi
n^{\mu },  \notag \\
\mathcal{R}^{\mu \nu }& =\varphi _{6}\Pi \pi ^{\mu \nu }+\varphi _{7}\pi
^{\lambda \left\langle \mu \right. }\pi _{\lambda }^{\left. \nu
\right\rangle }+\varphi _{8}n^{\left\langle \mu \right. }n^{\left. \nu
\right\rangle }.
\end{align}%
In Eq.\ (\ref{Final}), terms of order $\mathcal{O}(\mathrm{Kn}^{3})$, $%
\mathcal{O}(\mathrm{R}_{i}^{-1}\mathrm{R}_{j}^{-1}\mathrm{R}_{k}^{-1})$, $%
\mathcal{O}(\mathrm{Kn}^{2}\mathrm{R}_{i}^{-1})$ and $\mathcal{O}(\mathrm{Kn}%
\,\mathrm{R}_{i}^{-1}\mathrm{R}_{j}^{-1})$ were omitted.

Note that we have obtained equations of motion which are closed in terms of
14 dynamical variables. We remark that this was accomplished without making
use of the 14-moment approximation. This means that the reduction of degrees
of freedom was not obtained by a direct truncation of the moment expansion,
but by a separation of the microscopic time scales and the power-counting
scheme itself. The information about all other moments are actually included
in the transport coefficients, as will be shown later. If we also neglect
the terms of second order in inverse Reynolds number we recover the
equations of motion that are of the same form as those derived via the
14-moment approximation \cite{dkr}. However, even in this case, the
coefficients in Eqs.\ (\ref{14_moment_terms}) and relaxation times are not
the same as those calculated from the 14-moment approximation of Israel and
Stewart.

The resulting equations of motion (\ref{Final}) contain a large number of
transport coefficients. In particular, the viscosity coefficients and
relaxation times of the dissipative currents were found to be, 
\begin{align}
\tau _{\Pi }& =\frac{1}{\chi _{0}^{\left( 0\right) }},\text{ \ \ }\tau _{n}=%
\frac{1}{\chi _{0}^{\left( 1\right) }},\text{ \ \ }\tau _{\pi }=\frac{1}{%
\chi _{0}^{\left( 2\right) }},  \notag \\
\zeta & =\frac{m^{2}}{3}\sum_{r=0,\neq 1,2}^{N_{0}}\tau _{0r}^{\left(
0\right) }\alpha _{r}^{\left( 0\right) },\text{ \ \ }\kappa
_{n}=\sum_{r=0,\neq 1}^{N_{1}}\tau _{0r}^{\left( 1\right) }\alpha
_{r}^{\left( 1\right) },\text{ \ \ }\eta =\sum_{r=0}^{N_{2}}\tau
_{0r}^{\left( 2\right) }\alpha _{r}^{\left( 2\right) }.  \label{Results1}
\end{align}%
Note that in general these transport coefficients depend not only on one
moment of the distribution function but on all moments of corresponding rank 
$\ell $. As in Chapman-Enskog theory, the viscosity coefficients can only be
obtained by inverting $\mathcal{A}^{\left( \ell \right) }$. However, to
obtain the transient dynamics of the fluid, characterized by the relaxation
times, it is also necessary to find the eigenvalues and eigenvectors of $%
\mathcal{A}^{\left( \ell \right) }$.

In practice, the expansion (\ref{expansion2}) is always truncated at some
point and the matrices $\mathcal{A}^{\left( \ell \right) }$, $\Omega
^{\left( \ell \right) }$, and $\tau ^{\left( \ell \right) }$ will actually
be finite. The truncation of this expansion was already introduced as an
upper limit, $N_{\ell }$, in the corresponding summations. In principle, one
should only truncate the expansion (\ref{expansion2}) when the values of all
relevant transport coefficients have converged. Note that different
transport coefficients may require a different number of moments to converge.

\section{Applications}

\label{Applications}

In this section, we compute the transport coefficients for several cases.
First, we considered the lowest possible truncation scheme for Eq.\ (\ref%
{expansion2}) with $N_{0}=2$, $N_{1}=1$, and $N_{2}=0$. In this case, the
distribution function is expanded in terms of 14 moments and is actually
equivalent to the one obtained via Israel-Stewart's 14-moment ansatz.
Second, we consider the next simplest case and take $N_{0}=3$, $N_{1}=2$,
and $N_{2}=1$. Then, the distribution function is characterized by 23
moments, and consequently we shall refer to this case as 23-moment
approximation. Finally, we include 32 and 41 moments and verify that the
numerical values for the transport coefficients converge.

We also compute the transport coefficients of the terms appearing in $%
\mathcal{J}$, $\mathcal{J}^{\mu}$, and $\mathcal{J}^{\mu\nu}$ which are
displayed in Appendix \ref{transport coefficients}. These transport
coefficients were also calculated in previous derivations of fluid dynamics
from the Boltzmann equation. We shall explicitly point out the corrections
to the previous results introduced by our novel approach. Note, however,
that we are using a linear approximation to the collision term. Nonlinear
contributions could in principle also enter the transport coefficients in
the equations of motion (\ref{Final}), but will not be calculated here. Such
an investigation will be left for future work. For this reason we also do
not compute any coefficient of the terms of $\mathcal{O}(\mathrm{R}_{i}^{-1}%
\mathrm{R}_{j}^{-1})$, i.e., entering $\mathcal{R},\, \mathcal{R}^\mu$, and $%
\mathcal{R}^{\mu \nu}$, since all of them originate exclusively from
nonlinear contributions to the collision term.

\subsection{14-moment approximation}

The 14-moment approximation is recovered by truncating Eq.\ (\ref{expansion2}%
) at $N_{0}=2$, $N_{1}=1$, and $N_{2}=0$. For this specific truncation $%
\mathcal{A}^{(\ell )}$ is nothing but a number (because for $\mathcal{A}%
^{(0)}$ we have to exclude the second and third rows and columns and for $%
\mathcal{A}^{(1)}$ the second row and column), and thus 
\begin{equation*}
\tau ^{\left( \ell \right) }=\frac{1}{\mathcal{A}^{\left( \ell \right) }},%
\text{ \ \ }\Omega ^{\left( \ell \right) }=1,\text{ \ \ }\chi ^{\left( \ell
\right) }=\mathcal{A}^{\left( \ell \right) }.
\end{equation*}%
Then, the equations of motion and transport coefficients reduce to those
derived in Ref.\ \cite{dkr}.

For a classical gas of hard spheres with total cross section $\sigma $, in
the massless limit, the integrals $\mathcal{A}^{\left( 1\right) }=\mathcal{A}%
_{00}^{\left( 1\right) }$ and $\mathcal{A}^{\left( 2\right) }=\mathcal{A}%
_{00}^{\left( 2\right) }$ can be computed and have the simple form%
\begin{align}
\mathcal{A}^{\left( 1\right) }& =\frac{4}{9\lambda _{\mathrm{mfp}}}\;, \\
\mathcal{A}^{\left( 2\right) }& =\frac{3}{5\lambda _{\mathrm{mfp}}}\;,
\end{align}%
where we defined the mean free-path $\lambda _{\mathrm{mfp}}=1/\left(
n_{0}\sigma \right) $. The details of this calculation are shown in Appendix %
\ref{relax_times}. The coefficients in the ultra-relativistic limit, $m\beta
_{0}\rightarrow 0$, can then be calculated analytically. The coefficients of
order $\mathcal{O}(\mathrm{Kn}\,\mathrm{R}_{i}^{-1})$ are collected for the
shear stress and particle diffusion in Tables \ref{diff_massless} and \ref%
{shear_massless}. Note that, in this limit, the bulk viscous pressure
vanishes, and thus we do not need to compute $\mathcal{A}_{00}^{(0)}$. 
\begin{table}[h]
\begin{center}
\begin{tabular}{|c|c|c|c|c|c|c|}
\hline
$\kappa _{n}$ & $\tau _{n}[\lambda _{\mathrm{mfp}}]$ & $\delta _{nn}[\tau
_{n}]$ & ${\lambda }_{nn}[\tau _{n}]$ & ${\lambda }_{n\pi }[\tau _{n}]$ & $%
\ell _{n\pi }[\tau _{n}]$ & $\tau _{n\pi }[\tau _{n}]$ \\ \hline
${3}/\left( 16{\sigma }\right) $ & $9/4$ & $1$ & $3/5$ & $\beta _{0}/{20}$ & 
${\beta _{0}}/{20}$ & $0$ \\ \hline
\end{tabular}%
\end{center}
\caption{{\protect\small The coefficients for the particle diffusion for a
classical gas with constant cross section in the ultrarelativistic limit, in
the 14-moment approximation.}}
\label{diff_massless}
\end{table}
\begin{table}[h]
\begin{center}
\begin{tabular}{|c|c|c|c|c|c|c|}
\hline
$\eta $ & $\tau _{\pi }[\lambda _{\mathrm{mfp}}]$ & ${\tau }_{\pi \pi }[\tau
_{\pi}]$ & ${\lambda }_{\pi n}[\tau _{\pi}]$ & $\delta _{\pi \pi }[\tau
_{\pi}] $ & $\ell _{\pi n}[\tau _{\pi}]$ & $\tau _{\pi n}[\tau _{\pi}]$ \\ 
\hline
${4}/({3\sigma \beta _{0}})$ & $5/3$ & $10/7$ & $0$ & $4/3$ & $0$ & $0$ \\ 
\hline
\end{tabular}%
\end{center}
\caption{{\protect\small The coefficients for the shear stress for a
classical gas with constant cross section in the ultrarelativistic limit, in
the 14-moment approximation.}}
\label{shear_massless}
\end{table}

\subsection{Next correction: 23-moment approximation and beyond}

In order to better understand our formulas, Eqs.\ (\ref{Results1}), we would
like to compute the first correction to the expressions in Tables \ref%
{diff_massless} and \ref{shear_massless}. For this purpose, we consider $%
N_{0}=3$, $N_{1}=2$, and $N_{2}=1$. Then, $\mathcal{A}^{\left( \ell \right)
} $, $\Omega ^{\left( \ell \right) }$, and $\tau ^{\left( \ell \right) }$
are, after removing trivial rows and colums, $2\times 2$ matrices that can
be computed from the collision integral Eq.\ (\ref{integrals}). We obtain
the elements of $\mathcal{A}^{\left( 1,2\right) }$, its inverse $\tau
^{\left( 1,2\right) }$, and $\Omega ^{\left( 1,2\right) }$ as 
\begin{align}
\mathcal{A}^{\left( 1\right) }& =\frac{1}{3\lambda _{\mathrm{mfp}}}\left( 
\begin{array}{cc}
2 & \beta _{0}^{2}/30 \\ 
-4\beta _{0}^{-2} & 1%
\end{array}%
\right) ,\;\;\;\mathcal{A}^{\left( 2\right) }=\frac{1}{\lambda _{\mathrm{mfp}%
}}\left( 
\begin{array}{cc}
9/10 & -\beta _{0}/20 \\ 
4/\left( 3\beta _{0}\right) & 1/3%
\end{array}%
\right) ,  \label{matrix1} \\
\text{ }\tau ^{\left( 1\right) }& =\frac{3}{8}\lambda _{\mathrm{mfp}}\left( 
\begin{array}{cc}
15/4 & -\beta _{0}^{2}/8 \\ 
15\beta _{0}^{-2} & 15/2%
\end{array}%
\right) ,\;\;\;\tau ^{\left( 2\right) }=\frac{1}{11}\lambda _{\mathrm{mfp}%
}\left( 
\begin{array}{cc}
10 & 3\beta _{0}/2 \\ 
-40\beta _{0}^{-1} & 27%
\end{array}%
\right) , \\
\Omega ^{\left( 1\right) }& =\left( 
\begin{array}{cc}
1 & 1 \\ 
-\left( 15+\sqrt{105}\right) \beta _{0}^{-2} & \left( -15+\sqrt{105}\right)
\beta _{0}^{-2}%
\end{array}%
\right) ,\;\;\;\Omega ^{\left( 2\right) }=\left( 
\begin{array}{cc}
1 & 1 \\ 
8\beta _{0}^{-1} & 10/3\beta _{0}^{-1}%
\end{array}%
\right) ,  \label{matrix2}
\end{align}%
see Appendix \ref{relax_times} for details. The eigenvectors of $\mathcal{A}%
^{\left( 1\right) }$ and $\mathcal{A}^{\left( 2\right) }$ are%
\begin{align}
\chi _{0}^{\left( 1\right) }& =\frac{1}{2\lambda _{\mathrm{mfp}}}\left( 1-%
\sqrt{\frac{7}{135}}\right) \text{ },\text{ \ \ }\chi _{1}^{\left( 1\right)
}=\frac{1}{2\lambda _{\mathrm{mfp}}}\left( 1+\sqrt{\frac{7}{135}}\right) 
\text{ }, \\
\chi _{0}^{\left( 2\right) }& =\frac{1}{2\lambda _{\mathrm{mfp}}}\text{ },%
\text{ \ \ }\chi _{1}^{\left( 2\right) }=\frac{11}{15\lambda _{\mathrm{mfp}}}%
\text{ }.
\end{align}

Using the formulas derived in this paper, Eqs.~(\ref{Results1}), we
calculate the corrected values for the particle-number diffusion coefficient
and diffusion-relaxation time and for the shear viscosity and
shear-relaxation time,%
\begin{align}
\kappa _{n}& =\frac{21}{128}n_{0}\lambda _{\mathrm{mfp}}\simeq 0.164\text{ }%
n_{0}\lambda _{\mathrm{mfp}},  \label{Result1} \\
\tau _{n}& =\frac{90}{45-\sqrt{105}}\lambda _{\mathrm{mfp}}\simeq 2.5897%
\text{ }\lambda _{\mathrm{mfp}}\text{ },  \label{Result2} \\
\eta & =\frac{14}{11}P_{0}\lambda _{\mathrm{mfp}}\simeq 1.2727\text{ }%
P_{0}\lambda _{\mathrm{mfp}}\text{ },  \label{Result3} \\
\tau _{\pi }& =2\lambda _{\mathrm{mfp}}\text{ },  \label{Result4}
\end{align}%
where we used that, in the massless and classical limits,%
\begin{align*}
\alpha _{0}^{\left( 1\right) }& =\frac{1}{12}n_{0},\text{ \ \ \ \ }\alpha
_{2}^{\left( 1\right) }=-\frac{1}{\beta _{0}}P_{0}, \\
\alpha _{0}^{\left( 2\right) }& =\frac{4}{5}P_{0},\text{ \ \ \ \ \ }\alpha
_{1}^{\left( 2\right) }=\frac{4}{\beta _{0}}P_{0}.
\end{align*}%
As before, the coefficients in the ultra-relativistic limit, $m\beta
_{0}\rightarrow 0$, can then be calculated analytically. The coefficients of
order $\mathcal{O}(\mathrm{Kn}\,\mathrm{R}_{i}^{-1})$ are collected for the
shear stress and particle diffusion in Tables \ref{diff_massless2} and \ref%
{shear_massless2}. 
\begin{table}[h]
\begin{center}
\begin{tabular}{|c|c|c|c|c|c|c|}
\hline
$\kappa _{n}$ & $\tau _{n}[\lambda _{\mathrm{mfp}}]$ & $\delta _{nn}[\tau
_{n}]$ & $\lambda _{nn}[\tau _{n}]$ & $\lambda _{n\pi }[\tau _{n}]$ & $\ell
_{n\pi }[\tau _{n}]$ & $\tau _{n\pi }[\tau _{n}]$ \\ \hline
$21/\left( 128\sigma \right) $ & $2.59$ & $1.00$ & $0.96$ & $0.054\beta _{0}$
& $0.118\beta _{0}$ & $0.0295\beta _{0}/P_{0}$ \\ \hline
\end{tabular}%
\end{center}
\caption{{\protect\small The coefficients for the particle diffusion for a
classical gas with constant cross section in the ultrarelativistic limit, in
the 23-moment approximation.}}
\label{diff_massless2}
\end{table}
\begin{table}[h]
\begin{center}
\begin{tabular}{|c|c|c|c|c|c|c|}
\hline
$\eta $ & $\tau _{\pi }[\lambda _{\mathrm{mfp}}]$ & $\tau _{\pi \pi }[\tau
_{\pi}]$ & $\lambda _{\pi n}[\tau _{\pi}]$ & $\delta _{\pi \pi }[\tau
_{\pi}] $ & $\ell _{\pi n}[\tau _{\pi}]$ & $\tau _{\pi n}[\tau _{\pi}]$ \\ 
\hline
$14/(11\sigma \beta _{0})$ & $2$ & $134/77$ & $0.344\beta _{0}^{-1}$ & $4/3$
& $-0.689\beta _{0}^{-1}$ & $-0.689/n_{0}$ \\ \hline
\end{tabular}%
\end{center}
\caption{{\protect\small The coefficients for the shear stress for a
classical gas with constant cross section in the ultrarelativistic limit, in
the 23-moment approximation.}}
\label{shear_massless2}
\end{table}

To obtain these expressions we used the results from Appendix \ref{therm}
and that, in the massless/classical limits, $D_{20}=3P_{0}^{2}$. Note that
most of the transport coefficients were corrected by the inclusion of more
moments in the computation. The coefficients related to the shear-stress
tensor were less affected by the additional moments, when compared to the
particle-diffusion coefficients. This might explain the poor agreement
between the Israel-Stewart theory and numerical solutions of the Boltzmann
equation in Refs.\ \cite{BAMPS} regarding heat flow and fugacity.

We further checked the convergence of this approach by taking 32 and 41
moments. In this case, the matrices $\mathcal{A}^{\left( 1,2\right) }$, $%
\tau ^{\left( 1,2\right) }$\ and $\Omega ^{\left( 1,2\right) }$\ were
computed numerically. There is a clear tendency of convergence as we
increase the number of moments. For the particular case of classical
particles with constant cross sections, 32 moments seems sufficient. See
Tables \ref{diff_massless3} and \ref{shear_massless33} for the results. 
\begin{table}[h]
\begin{center}
\begin{tabular}{|c|c|c|c|c|c|c|c|}
\hline
number of moments & $\kappa _{n}$ & $\tau _{n}[\lambda _{\mathrm{mfp}}]$ & $%
\delta _{nn}[\tau _{n }]$ & $\lambda _{nn}[\tau _{n }]$ & $\lambda _{n\pi
}[\tau _{n }]$ & $\ell _{n\pi }[\tau _{n }]$ & $\tau _{n\pi }[\tau _{n }]$
\\ \hline
$14$ & ${3}/\left( 16{\sigma }\right) $ & $9/{4}$ & $1$ & $3/5$ & $\beta
_{0}/{20}$ & ${\beta _{0}}/{20}$ & $0$ \\ \hline
$23$ & $21/\left( 128\sigma \right) $ & $2.59$ & $1.0$ & $0.96$ & $%
0.054\beta _{0}$ & $0.118\beta _{0}$ & $0.0295\beta _{0}/P_{0}$ \\ \hline
$32$ & $0.1605/\sigma $ & $2.57$ & $1.0$ & $0.93$ & $0.052\beta _{0}$ & $%
0.119\beta _{0}$ & $0.0297\beta _{0}/P_{0}$ \\ \hline
$41$ & $0.1596/\sigma $ & $2.57$ & $1.0$ & $0.92$ & $0.052\beta _{0}$ & $%
0.119\beta _{0}$ & $0.0297\beta _{0}/P_{0}$ \\ \hline
\end{tabular}%
\end{center}
\caption{{\protect\small The coefficients for the particle diffusion for a
classical gas with constant cross section in the ultrarelativistic limit, in
the 14, 23, 32 and 41-moment approximation.}}
\label{diff_massless3}
\end{table}
\begin{table}[h]
\begin{center}
\begin{tabular}{|c|c|c|c|c|c|c|c|}
\hline
number of moments & $\eta $ & $\tau _{\pi }[\lambda _{\mathrm{mfp}}]$ & $%
\tau _{\pi \pi }[\tau _{\pi }] $ & $\lambda _{\pi n}[\tau _{\pi }]$ & $%
\delta _{\pi \pi }[\tau _{\pi }]$ & $\ell _{\pi n}[\tau _{\pi }] $ & $\tau
_{\pi n}[\tau _{\pi }] $ \\ \hline
$14$ & ${4}/({3\sigma \beta _{0}})$ & ${5}/3$ & $10/7$ & $0$ & $4/3$ & $0$ & 
$0$ \\ \hline
$23$ & $14/(11\sigma \beta _{0})$ & $2$ & $134/77$ & $0.344\beta _{0}^{-1}$
& $4/3$ & $-0.689/\beta _{0}$ & $-0.689/n_{0}$ \\ \hline
$32$ & $1.268/(\sigma \beta _{0})$ & $2$ & $1.69$ & $0.254\beta _{0}^{-1}$ & 
$4/3$ & $-0.687/\beta _{0}$ & $-0.687/n_{0}$ \\ \hline
$41$ & $1.267/(\sigma \beta _{0})$ & $2$ & $1.69$ & $0.244\beta _{0}^{-1}$ & 
$4/3$ & $-0.685/\beta _{0}$ & $-0.685/n_{0}$ \\ \hline
\end{tabular}%
\end{center}
\caption{{\protect\small The coefficients for the shear stress for a
classical gas with constant cross section in the ultrarelativistic limit, in
the 14, 23, 32 and 41-moment approximation.}}
\label{shear_massless33}
\end{table}

\section{Discussion and Conclusions}

\label{conclusions}

\subsection{Knudsen number and the reduction of dynamical variables}

It is important to mention that the terms $\mathcal{K},\,\mathcal{K}^{\mu }$%
, and $\mathcal{K}^{\mu \nu }$ which are of second order in Knudsen number
lead to several problems. The terms which contain second-order spatial
derivatives of $u^{\mu }$, $\alpha _{0}$, and $P_{0}$, e.g., $\nabla _{\mu
}I^{\mu }$, $\nabla _{\mu }F^{\mu }$, $\nabla ^{\left\langle \mu \right.
}I^{\left. \nu \right\rangle }$, $\nabla ^{\left\langle \mu \right.
}F^{\left. \nu \right\rangle }$, $\Delta _{\alpha }^{\mu }\partial _{\nu
}\sigma ^{\alpha \nu }$, and $\nabla ^{\mu }\theta $, are especially
problematic since they change the boundary conditions of the equations. In
relativistic systems these derivatives, even though they are space-like,
also contain time derivatives and thus require initial values. This means
that, by including them, one would have to specify not only the initial
spatial distribution of the fluid-dynamical variables but also the spatial
distribution of their time derivatives. In practice, this implies that we
would be increasing the number of fluid-dynamical degrees of freedom.

There is an even more serious problem. By including terms of order higher
than one in Knudsen number, the transport equations become parabolic. In a
relativistic theory, this comes with disastrous consequences since the
solutions are acausal and consequently unstable \cite{his}. For this reason,
if one wants to include terms of higher order in Knudsen number, it is
mandatory to include also second-order co-moving time derivatives of the
dissipative quantities. Or, equivalently, one could promote the moments $%
\rho_3,\,\rho_2^\mu, \rho_1^{\mu \nu}$ or further ones to dynamical
variables. For this reason we do not compute the transport coefficients for
these higher-order terms in this paper.

In practice, a way around this would be to replace e.g.\ the $%
\sigma^{\lambda \langle \mu} \sigma_{\lambda}^{\nu\rangle}$ term in $%
\mathcal{K}^{\mu \nu}$ using the asymptotic (Navier-Stokes) solution by $%
(1/2\eta)\pi^{\lambda \langle \mu} \sigma_{\lambda}^{\nu\rangle}$, and thus
effectively rendering it a term contributing to $\mathcal{\ J}^{\mu \nu}$.
This should be a reasonable approximation if one is sufficiently close to
the asymptotic solution. This would then change the coefficient of the
respective term in $\mathcal{J}^{\mu \nu}$. In principle, this could be done
to all terms in $\mathcal{K},\, \mathcal{K}^\mu,$ and $\mathcal{K}^{\mu \nu}$%
, except for the ones containing exclusively powers and/or gradients of $%
F^\mu$ and $\omega^{\mu \nu}$. In the same spirit, using the asymptotic
solutions one could also shuffle some of the terms in $\mathcal{J},\, 
\mathcal{J}^\mu$, and $\mathcal{J}^{\mu \nu}$ (those not containing $F^\mu$, 
$\omega^{\mu \nu}$, and gradients of dissipative currents) into terms
contributing to $\mathcal{R},\, \mathcal{R}^\mu$, and $\mathcal{R}^{\mu \nu}$
(or vice versa). How this changes the actual transient dynamics remains to
be investigated in the future.

\subsection{Navier-Stokes limit}

Note that one of the main features of transient theories of fluid dynamics
is the relaxation of the dissipative currents towards their Navier-Stokes
values, on time scales given by the transport coefficients $\tau _{\Pi }$, $%
\tau _{n}$, and $\tau _{\pi }$. From the Boltzmann equation, Navier-Stokes
theory is obtained by means of the Chapman-Enskog expansion which describes
an asymptotic solution of the single-particle distribution function. It is
already clear from the previous section that the equations of motion derived
in this paper approach Navier-Stokes-type solutions at asymptotically long
times, in which the dissipative currents are solely expressed in terms of
gradients of fluid-dynamical variables.

It is interesting to investigate, however, if our equations approach the
correct Navier-Stokes theory, i.e., if the viscosity coefficients obtained
via our method are equivalent to the ones obtained via Chapman-Enskog
theory. It should be noted that this is not the case for Grad's and Israel
and Stewart's theories \cite{IS, DeGroot,dkr}. The viscosity coefficients
computed by these theories do not coincide with those extracted from the
Chapman-Enskog theory. We remark that, after taking into account the first
corrections to the shear viscosity coefficient, see Eq.\ (\ref{Result3}) and
Table \ref{shear_massless33}, our result approached the solution obtained
using Chapman-Enskog theory, $\eta_{NS}=1.2654/\left( \beta_{0}\sigma\right) 
$ \cite{DeGroot}. In principle there is no reason for the method of moments
to attain a different Navier-Stokes limit than Chapman-Enskog theory. We can
show that, if the same basis of irreducible tensors $k^{\langle \mu_1}
\cdots k^{\mu_\ell \rangle}$ and polynomials $P_{n\mathbf{k}}^{(\ell)}$ is
used in both calculations, they both yield the same result, even order by
order.

\subsection{ \textquotedblleft Non-hydrodynamic\textquotedblright\ modes and
the microscopic origin of the relaxation time}

One of the features of the theory derived in this paper (and also of Grad's
and Israel-Stewart's theories) is the appearance of so-called
non-hydrodynamics modes, i.e., modes that do not vanish in the limit of zero
wave-number. Such modes do not exist in Navier-Stokes theory or its
extensions via the Chapman-Enskog expansion. For this reason, these modes
are usually not associated with fluid-dynamical behavior, hence the label
"non-hydrodynamic".

The non-hydrodynamic modes describe the relaxation of the dissipative
currents towards their respective Navier-Stokes solutions and can be
directly related to the respective relaxation times. For the case of the
shear non-hydrodynamic mode, $\omega _{\mathrm{shear}}^{\mathrm{non-hydro}%
}\left( \mathbf{k}\right) $, it can be shown that in the limit of $\mathbf{k}%
\rightarrow 0$ the mode is given by $\omega _{\mathrm{shear}}^{\mathrm{%
non-hydro}}\left( \mathbf{0}\right) =-i/\tau _{\pi }$ \cite{his}. In
Chapman-Enskog theory the transient dynamics of the system is neglected,
e.g., it is assumed that in the absence of space-like gradients, time-like
gradients vanish as well, and it is natural that such modes do not exist.

The appearance of non-hydrodynamic modes in a fluid-dynamical theory seems
to counteract the prevalent belief that fluid dynamics effectively describes
the asymptotic long-time and long-distance behavior of the microscopic
theory. Recently, a microscopic formula for the relaxation time of
dissipative currents was obtained in the framework of linear response theory 
\cite{paperpoles}. In that paper, the relaxation time was shown to be
intrinsically related to the slowest microscopic time scale of the system,
i.e., to the singularity of the retarded Green's function closest to the
origin in the complex-plane. Thus, the non-hydrodynamic modes in Israel and
Stewart's theory and in the equations derived in this paper belong to a
description at long, but not asymptotically long, times.

This means that the theory derived in this paper (as well as Israel and
Stewart's theory) attempts to describe the dynamics of the dissipative
currents at time scales of the order of the (slowest) microscopic times
scale (which is of the order of the mean-free path). Such findings challenge
the point of view that a fluid-dynamical description can only be formulated
around zero frequency and wave number and that the inclusion of relaxation
times can only be understood as a regularization method to control the
instabilities of the gradient expansion. In fact, the relaxation times
correspond to microscopic time scales, independent of any macroscopic scale
related to the gradients of fluid-dynamical variables. Note that the
expressions presented in Ref.\ \cite{paperpoles} and in this paper for $\eta 
$ and $\tau _{\pi }$ are equivalent.

\subsection{Conclusions}

In this work we have presented a general and consistent derivation of
relativistic fluid dynamics from the Boltzmann equation using the method of
moments. First, a general expansion of the single-particle distribution
function in terms of its moments was introduced in Sec.\ \ref{Mom_Meth}. We
constructed an orthonormal basis which allowed us to expand and obtain exact
relations between the expansion parameters and irreducible moments of the
deviations of the distribution function from equilibrium. We then proceeded
to derive exact equations for these moments.

The main difference of our approach to previous work is that we did not
close the fluid-dynamical equations of motion by truncating the expansion of
the distribution function. Instead, we kept all terms in the moment
expansion and truncated the exact equations of motion according to a
power-counting scheme in Knudsen and inverse Reynolds number. Contrary to
many calculations, we did not assume that the inverse Reynolds and Knudsen
numbers are of the same order. As a matter of fact, in order to obtain
relaxation-type equations, we had to explicitly include the slowest
microscopic time scales, which are shown to be the characteristic times
within which dissipative currents relax towards their asymptotic
Navier-Stokes solutions. Thus, Navier-Stokes theory, or the Chapman-Enskog
expansion, is already included in our formulation as an asymptotic limit of
the dynamical equations derived in this paper.

We concluded that the equations of motion can be closed in terms of only 14
dynamical variables, as long as we only keep terms of second order in
Knudsen and/or inverse Reynolds number. Even though the equations of motion
are closed in terms of these 14 fields, the transport coefficients carry
information about all moments of the distribution function (all the
different relaxation scales of the irreducible moments). The bulk-viscosity,
particle-diffusion, and shear-viscosity coefficients agree with the values
obtained via Chapman-Enskog theory.

\section{Acknowledgments}

G.S.D.\ and H.N.\ acknowledge the hospitality of MTA-KFKI, Budapest, where
part of this work was accomplished. G.S.D and D.H.R. acknowledge the
hospitality of the High-Energy Physics group at Jyv\"{a}skyl\"{a} University
where this work was completed. The authors thank T.\ Koide for enlightening
discussions. This work was supported by the Helmholtz International Center
for FAIR within the framework of the LOEWE program launched by the State of
Hesse. The work of H.N.\ was supported by the Extreme Matter Institute
(EMMI). E.M.\ was supported by Hungarian National Development Agency OTKA/NF%
\"{U} 81655.

\appendix

\section{Derivation of the collision terms}

\label{collision}

In this appendix, we derive Eqs.\ (\ref{great_formula}) and (\ref{integrals}%
). The first step is to linearize the collision operator, 
\begin{equation}
C\left[ f\right] =\frac{1}{\nu }\int dK^{\prime }dPdP^{\prime }W_{\mathbf{kk}%
\prime \rightarrow \mathbf{pp}\prime }\left( f_{\mathbf{p}}f_{\mathbf{p}%
^{\prime }}\tilde{f}_{\mathbf{k}}\tilde{f}_{\mathbf{k}^{\prime }}-f_{\mathbf{%
k}}f_{\mathbf{k}^{\prime }}\tilde{f}_{\mathbf{p}}\tilde{f}_{\mathbf{p}%
^{\prime }}\right) ,  \label{A0}
\end{equation}%
in the deviations from the equilibrium distribution functions. In the main
text, the deviations from the local-equilibrium distribution function were
parametrized as 
\begin{equation}
\delta f_{\mathbf{p}}=f_{\mathbf{p}}-f_{0\mathbf{p}}=f_{0\mathbf{p}}\tilde{f}%
_{0\mathbf{p}}\phi _{\mathbf{p}}\;.
\end{equation}%
Then, only keeping terms of first order in $\phi $, we can prove that%
\begin{eqnarray}
f_{\mathbf{p}}f_{\mathbf{p}^{\prime }} &=&f_{0\mathbf{p}}f_{0\mathbf{p}%
^{\prime }}\left( 1+\tilde{f}_{0\mathbf{p}^{\prime }}\phi _{\mathbf{p}%
^{\prime }}+\tilde{f}_{0\mathbf{p}}\phi _{\mathbf{p}}\right) +\mathcal{O}%
\left( \phi ^{2}\right) ,  \label{A1} \\
\tilde{f}_{\mathbf{p}}\tilde{f}_{\mathbf{p}^{\prime }} &=&\tilde{f}_{0%
\mathbf{p}}\tilde{f}_{0\mathbf{p}^{\prime }}\left( 1-af_{0\mathbf{p}^{\prime
}}\phi _{\mathbf{p}^{\prime }}-af_{0\mathbf{p}}\phi _{\mathbf{p}}\right) +%
\mathcal{O}\left( \phi ^{2}\right) \;.  \label{A2}
\end{eqnarray}%
Substituting Eqs.\ (\ref{A1}) and (\ref{A2}) into Eq.\ (\ref{A0}), we
obtain, 
\begin{equation}
C\left[ f\right] =\frac{1}{\nu }\int dK^{\prime }dPdP^{\prime }W_{\mathbf{kk}%
\prime \rightarrow \mathbf{pp}\prime }f_{0\mathbf{k}}f_{0\mathbf{k}^{\prime
}}\tilde{f}_{0\mathbf{p}}\tilde{f}_{0\mathbf{p}^{\prime }}\left( \phi _{%
\mathbf{p}}+\phi _{\mathbf{p}^{\prime }}-\phi _{\mathbf{k}}-\phi _{\mathbf{k}%
^{\prime }}\right) +\mathcal{O}\left( \phi ^{2}\right) \;,  \label{Lin_Col}
\end{equation}%
where we also used the equalities 
\begin{eqnarray}
\tilde{f}_{0\mathbf{p}} &=&f_{0\mathbf{p}}\exp \left( \beta _{0}E_{\mathbf{p}%
}-\alpha _{0}\right) \;, \\
f_{0\mathbf{p}}f_{0\mathbf{p}^{\prime }}\tilde{f}_{0\mathbf{k}}\tilde{f}_{0%
\mathbf{k}^{\prime }} &=&f_{0\mathbf{k}}f_{0\mathbf{k}^{\prime }}\tilde{f}_{0%
\mathbf{p}}\tilde{f}_{0\mathbf{p}^{\prime }}\;.
\end{eqnarray}%
Inserting Eq.\ (\ref{Lin_Col}) in the expression for the irreducible
collision term (\ref{General_Col_term}), we obtain%
\begin{eqnarray}
C_{r-1}^{\left\langle \mu _{1}\cdots \mu _{\ell }\right\rangle } &=&\frac{1}{%
\nu }\int dKdK^{\prime }dPdP^{\prime }W_{\mathbf{kk}\prime \rightarrow 
\mathbf{pp}\prime }f_{0\mathbf{k}}f_{0\mathbf{k}^{\prime }}\tilde{f}_{0%
\mathbf{p}}\tilde{f}_{0\mathbf{p}^{\prime }}  \notag \\
&&\times E_{\mathbf{k}}^{r-1}k^{\left\langle \mu _{1}\right. }\cdots
k^{\left. \mu _{\ell }\right\rangle }\left( \phi _{\mathbf{p}}+\phi _{%
\mathbf{p}^{\prime }}-\phi _{\mathbf{k}}-\phi _{\mathbf{k}^{\prime }}\right)
+\mathcal{O}\left( \phi ^{2}\right) .  \label{asd}
\end{eqnarray}%
The next step is to substitute the moment expansion of the single-particle
distribution function, Eqs.\ (\ref{expansion1}) and (\ref{bla11}), into Eq.\
(\ref{asd}), expressing it in the following form%
\begin{equation}
C_{r-1}^{\left\langle \mu _{1}\cdots \mu _{\ell }\right\rangle
}=-\sum_{m=0}^{\infty }\sum_{n=0}^{N_{m}}\left( \mathcal{A}_{rn}\right)
_{\nu _{1}\cdots \nu _{m}}^{\mu _{1}\cdots \mu _{\ell }}\rho _{n}^{\nu
_{1}\cdots \nu _{m}}+\mathcal{O}\left( \phi ^{2}\right) ,  \label{Original}
\end{equation}%
where we defined the tensor%
\begin{eqnarray}
\left( \mathcal{A}_{rn}\right) _{\nu _{1}\cdots \nu _{m}}^{\mu _{1}\cdots
\mu _{\ell }} &\equiv &\frac{1}{\nu }\int dKdK^{\prime }dPdP^{\prime }W_{%
\mathbf{kk}\prime \rightarrow \mathbf{pp}\prime }f_{0\mathbf{k}}f_{0\mathbf{k%
}^{\prime }}\tilde{f}_{0\mathbf{p}}\tilde{f}_{0\mathbf{p}^{\prime }}E_{%
\mathbf{k}}^{r-1}k^{\left\langle \mu _{1}\right. }\cdots k^{\left. \mu
_{\ell }\right\rangle }  \notag \\
&&\times \left( \mathcal{H}_{\mathbf{k}n}^{\left( m\right)
}\,k_{\left\langle \nu _{1}\right. }\cdots k_{\left. \nu _{m}\right\rangle }+%
\mathcal{H}_{\mathbf{k}^{\prime }n}^{\left( m\right) }\,k_{\left\langle \nu
_{1}\right. }^{\prime }\cdots k_{\left. \nu _{m}\right\rangle }^{\prime }-%
\mathcal{H}_{\mathbf{p}n}^{\left( m\right) }\,p_{\left\langle \nu
_{1}\right. }\cdots p_{\left. \nu _{m}\right\rangle }-\mathcal{H}_{\mathbf{p}%
^{\prime }n}^{\left( m\right) }\,p_{\left\langle \nu _{1}\right. }^{\prime
}\cdots p_{\left. \nu _{m}\right\rangle }^{\prime }\right) .  \label{tensor}
\end{eqnarray}%
The integral $\left( \mathcal{A}_{rn}\right) _{\nu _{1}\cdots \nu _{m}}^{\mu
_{1}\cdots \mu _{\ell }}$ is a tensor of rank $m+\ell $, which is symmetric
under permutations of $\mu $--type indices and symmetric under permutations
of $\nu $--type indices, and which depends only on equilibrium distribution
functions. The latter contain only the fluid 4-velocity $u^{\mu }$.
Therefore, $\left( \mathcal{A}_{rn}\right) _{\nu _{1}\cdots \nu _{m}}^{\mu
_{1}\cdots \mu _{\ell }}$ must be constructed from tensor structures made of 
$u^{\mu }$ and the metric tensor $g^{\mu \nu }$. Also, $\left( \mathcal{A}%
_{rn}\right) _{\nu _{1}\cdots \nu _{m}}^{\mu _{1}\cdots \mu _{\ell }}$ was
constructed to be orthogonal to $u^{\mu }$ and to satisfy the following
property,%
\begin{equation}
\Delta _{\mu _{1}\cdots \mu _{\ell }}^{\alpha _{1}\cdots \alpha _{\ell
}}\Delta _{\beta _{1}\cdots \beta _{m}}^{\nu _{1}\cdots \nu _{m}}\left( 
\mathcal{A}_{rn}\right) _{\nu _{1}\cdots \nu _{m}}^{\mu _{1}\cdots \mu
_{\ell }}=\left( \mathcal{A}_{rn}\right) _{\beta _{1}\cdots \beta
_{m}}^{\alpha _{1}\cdots \alpha _{\ell }}.  \label{great property}
\end{equation}%
Since $\left( \mathcal{A}_{rn}\right) _{\nu _{1}\cdots \nu _{m}}^{\mu
_{1}\cdots \mu _{\ell }}$ is orthogonal to $u^{\mu }$, it can only be
constructed from combinations of projection operators, $\Delta ^{\mu \nu }$.
This already constrains $m+\ell $ to be an even number, since it is
impossible to construct odd-ranked tensors solely from $\Delta ^{\mu \nu }$%
s. This means that both $\ell $ and $m$ are either even or odd. Therefore,
the following type of terms could appear in $\left( \mathcal{A}_{rn}\right)
_{\nu _{1}\cdots \nu _{m}}^{\mu _{1}\cdots \mu _{\ell }}$:

\begin{enumerate}
\item[(i)] Terms where all $\mu $--type indices pair up on projectors $%
\Delta ^{\mu _{i}\mu _{j}}$ and all $\nu $--type indices on projectors $%
\Delta _{\nu _{p}\nu _{q}}$, e.g. 
\begin{equation}
\Delta ^{\mu _{1}\mu _{2}}\cdots \Delta ^{\mu _{i}\mu _{j}}\cdots \Delta
^{\mu _{\ell -1}\mu _{\ell }}\Delta _{\nu _{1}\nu _{2}}\cdots \Delta _{\nu
_{p}\nu _{q}}\cdots \Delta _{\nu _{m-1}\nu _{m}}\;.
\end{equation}%
All possible permutations of the $\mu $--type indices among themselves and $%
\nu $--type indices among themselves are allowed.

\item[(ii)] Terms where at least one $\mu$--type index pairs with a $\nu$%
--type index on a projector, e.g. 
\begin{equation}
\Delta^{\mu_1}_{\nu_1} \Delta^{\mu_2 \mu_3} \cdots \Delta^{\mu_i\mu_j}
\cdots \Delta^{\mu_{\ell-1}\mu_\ell}\Delta_{\nu_2\nu_3} \cdots
\Delta_{\nu_p\nu_q} \cdots \Delta_{\nu_{m-1}\nu_m}\;.
\end{equation}
Again, all possible permutations of the $\mu$--type and $\nu$--type indices
are allowed. If there is an odd number of projectors of the type $%
\Delta^{\mu_i}_{\nu_p}$, both $\ell$ and $m$ must be odd. If there is an
even number, both $\ell$ and $m$ must be even, too. Without loss of
generality, suppose that $\ell > m$. For $\ell +m$ to be even, $\ell$ must
be $m+2,m+4,\ldots$. Then one could pair all $\nu$--type indices with $\mu$%
--type indices on projectors of the form $\Delta^{\mu_i}_{\nu_p}$, with some
projectors left over which carry only $\mu$--type indices, e.g.\ $%
\Delta^{\mu_j \mu_k}$.

\item[(iii)] If $\ell = m$, all $\mu$--type indices could be paired up with $%
\nu$--type indices on projectors of the form $\Delta^{\mu_i}_{\nu_p}$, with
no left-over projectors like explained at the end of (ii), 
\begin{equation}
\Delta_{\nu_{1}}^{\mu_{1}}\cdots \Delta_{\nu_{\ell}}^{\mu_\ell}\;.
\end{equation}
Again, all permutations of the $\mu$--type indices among themselves and $\nu$%
--type indices among themselves are allowed.
\end{enumerate}

Note that terms of the type (i) and (ii) by themselves do not satisfy the
property (\ref{great property}). This happens because any term which
contains at least one projector of the type $\Delta ^{\mu _{i}\mu _{j}}$ or $%
\Delta _{\nu _{p}\nu _{q}}$ vanishes when contracted with $\Delta _{\mu
_{1}\cdots \mu _{\ell }}^{\alpha _{1}\cdots \alpha _{\ell }}\Delta _{\beta
_{1}\cdots \beta _{m}}^{\nu _{1}\cdots \nu _{m}}$. Therefore, $\left( 
\mathcal{A}_{rn}\right) _{\nu _{1}\cdots \nu _{m}}^{\mu _{1}\cdots \mu
_{\ell }}$ cannot be solely constructed from terms of type (i) and (ii),
because otherwise it would vanish trivially, and property (\ref{great
property}) would not be satisfied. There must at least be one term of type
(iii). However, this implies that $m=\ell $. This does not imply that terms
of type (i) and (ii) do not appear; they do occur, but in such a way that
Eq.\ (\ref{great property}) is satisfied. In summary, $\left( \mathcal{A}%
_{rn}\right) _{\nu _{1}\cdots \nu _{m}}^{\mu _{1}\cdots \mu _{\ell }}$ has
the form 
\begin{equation}
\left( \mathcal{A}_{rn}\right) _{\nu _{1}\cdots \nu _{m}}^{\mu _{1}\cdots
\mu _{\ell }}=\delta _{\ell m}\left\{ \mathcal{A}_{rn}^{\left( \ell \right)
}\Delta _{\left( \nu _{1}\right. }^{\left( \mu _{1}\right. }\cdots \Delta
_{\left. \nu _{\ell }\right) }^{\left. \mu _{\ell }\right) }+[%
\mbox{terms of type (i)
and (ii)}]\right\} \;,  \label{finaly}
\end{equation}%
where the parentheses denote the symmetrization of all Lorentz indices.
Contracting Eq.\ (\ref{finaly}) with $\Delta _{\mu _{1}\cdots \mu _{\ell
}}^{\alpha _{1}\cdots \alpha _{\ell }}\Delta _{\beta _{1}\cdots \beta _{\ell
}}^{\nu _{1}\cdots \nu _{\ell }}$ and using Eq.\ (\ref{great property}), we
prove that%
\begin{equation}
\left( \mathcal{A}_{rn}\right) _{\beta _{1}\cdots \beta _{m}}^{\alpha
_{1}\cdots \alpha _{\ell }}=\delta _{\ell m}\,\mathcal{A}_{rn}^{\left( \ell
\right) }\,\Delta _{\beta _{1}\cdots \beta _{\ell }}^{\alpha _{1}\cdots
\alpha _{\ell }}\;.  \label{nontrivial}
\end{equation}%
Finally, substituting Eq.\ (\ref{nontrivial}) into Eq.\ (\ref{Original}) we
derive Eq.\ (\ref{great_formula}), introduced in the main text of the paper, 
\begin{equation}
C_{r-1}^{\left\langle \mu _{1}\cdots \mu _{\ell }\right\rangle
}=-\sum_{m=0}^{\infty }\mathcal{A}_{rn}^{\left( \ell \right) }\rho _{n}^{\mu
_{1}\cdots \mu _{\ell }}\;.
\end{equation}%
The coefficients $\mathcal{A}_{rn}^{\left( \ell \right) }$ can be obtained
from the following projection of $\left( \mathcal{A}_{rn}\right) _{\nu
_{1}\cdots \nu _{\ell }}^{\mu _{1}\cdots \mu _{\ell }}$, 
\begin{eqnarray}
\mathcal{A}_{rn}^{\left( \ell \right) } &=&\frac{1}{\Delta _{\mu _{1}\cdots
\mu _{\ell }}^{\mu _{1}\cdots \mu _{\ell }}}\Delta _{\mu _{1}\cdots \mu
_{\ell }}^{\nu _{1}\cdots \nu _{\ell }}\left( \mathcal{A}_{rn}\right) _{\nu
_{1}\cdots \nu _{\ell }}^{\mu _{1}\cdots \mu _{\ell }},  \notag \\
&=&\frac{1}{\nu \left( 2\ell +1\right) }\int dKdK^{\prime }dPdP^{\prime }W_{%
\mathbf{kk}\prime \rightarrow \mathbf{pp}\prime }f_{0\mathbf{k}}f_{0\mathbf{k%
}^{\prime }}\tilde{f}_{0\mathbf{p}}\tilde{f}_{0\mathbf{p}^{\prime }}E_{%
\mathbf{k}}^{r-1}k^{\left\langle \mu _{1}\right. }\cdots k^{\left. \mu
_{\ell }\right\rangle }  \notag \\
&&\times \left( \mathcal{H}_{\mathbf{k}n}^{\left( \ell \right)
}\,k_{\left\langle \mu _{1}\right. }\cdots k_{\left. \mu _{\ell
}\right\rangle }+\mathcal{H}_{\mathbf{k}^{\prime }n}^{\left( \ell \right)
}\,k_{\left\langle \mu _{1}\right. }^{\prime }\cdots k_{\left. \mu _{\ell
}\right\rangle }^{\prime }-\mathcal{H}_{\mathbf{p}n}^{\left( \ell \right)
}\,p_{\left\langle \mu _{1}\right. }\cdots p_{\left. \mu _{\ell
}\right\rangle }-\mathcal{H}_{\mathbf{p}^{\prime }n}^{\left( \ell \right)
}\,p_{\left\langle \mu _{1}\right. }^{\prime }\cdots p_{\left. \mu _{\ell
}\right\rangle }^{\prime }\right) ,
\end{eqnarray}%
where we used that $\Delta _{\mu _{1}\cdots \mu _{\ell }}^{\mu _{1}\cdots
\mu _{\ell }}=2\ell +1$.

\section{Calculation of the collision integrals}

\label{relax_times}

In this appendix, we calculate the collision integrals, Eq.\ (\ref{integrals}%
), for a classical gas, i.e., $\tilde{f}_{0\mathbf{k}}=1$, of hard spheres
in the ultrarelativistic limit, $m\beta _{0}\ll 1$. Then, Eq.\ (\ref%
{integrals}) becomes%
\begin{align}
\mathcal{A}_{rn}^{\left( \ell \right) }& =\frac{1}{\nu \left( 2\ell
+1\right) }\int dKdK^{\prime }dPdP^{\prime }W_{\mathbf{kk}\prime \rightarrow 
\mathbf{pp}\prime }f_{0\mathbf{k}}f_{0\mathbf{k}\prime }  \notag \\
& \times E_{\mathbf{k}}^{r-1}k^{\left\langle \nu _{1}\right. }\cdots
k^{\left. \nu _{\ell }\right\rangle }\left( \mathcal{H}_{\mathbf{k}%
n}^{\left( \ell \right) }k_{\left\langle \nu _{1}\right. }\cdots k_{\left.
\nu _{\ell }\right\rangle }+\mathcal{H}_{\mathbf{k}^{\prime }n}^{\left( \ell
\right) }k_{\left\langle \nu _{1}\right. }^{\prime }\cdots k_{\left. \nu
_{\ell }\right\rangle }^{\prime }-\mathcal{H}_{\mathbf{p}n}^{\left( \ell
\right) }p_{\left\langle \nu _{1}\right. }\cdots p_{\left. \nu _{\ell
}\right\rangle }-\mathcal{H}_{\mathbf{p}^{\prime }n}^{\left( \ell \right)
}p_{\left\langle \nu _{1}\right. }^{\prime }\cdots p_{\left. \nu _{\ell
}\right\rangle }^{\prime }\right) \;.
\end{align}%
The functions $\mathcal{H}_{\mathbf{k}n}^{\left( \ell \right) }$ were
defined in the main text, see Eq.\ (\ref{Hk}). The transition rate $W_{%
\mathbf{kk}\prime \rightarrow \mathbf{pp}\prime }$ is written in terms of
the differential cross section $\sigma (s,\Theta )$ as%
\begin{equation}
W_{\mathbf{kk}\prime \rightarrow \mathbf{pp}\prime }=s\sigma (s,\Theta
)\left( 2\pi \right) ^{6}\delta ^{(4)}\left( k^{\mu }+k^{\prime \mu }-p^{\mu
}-p^{\prime \mu }\right) .
\end{equation}%
The variable $s$ and $\Theta $ are defined as%
\begin{equation}
s=\left( k+k^{\prime }\right) ^{2},\text{ \ \ \ }\cos \Theta =\frac{\left(
k-k^{\prime }\right) \cdot \left( p-p^{\prime }\right) }{\left( k-k^{\prime
}\right) ^{2}}.
\end{equation}%
We further define the total cross section as the integral%
\begin{equation}
\sigma _{T}(s)=\frac{2\pi }{\nu }\int d\Theta \,\sin \Theta \,\sigma
(s,\Theta )\;.
\end{equation}%
In order to calculate $\mathcal{A}_{rn}^{\left( \ell \right) }$ it is
convenient to first define the tensors $X_{\mu \nu \gamma _{1}\cdots \gamma
_{m}}^{n}$%
\begin{align}
X_{\mu \nu \gamma _{1}\cdots \gamma _{m}}^{n}& =\frac{1}{\nu }\int
dKdK^{\prime }dPdP^{\prime }W_{\mathbf{kk}\prime \rightarrow \mathbf{pp}%
\prime }f_{0\mathbf{k}}f_{0\mathbf{k}^{\prime }}  \notag \\
& \times E_{\mathbf{k}}^{n}k_{\mu }k_{\nu }\left( k_{\gamma _{1}}\cdots
k_{\gamma _{m}}+k_{\gamma _{1}}^{\prime }\cdots k_{\gamma _{m}}^{\prime
}-p_{\gamma _{1}}\cdots p_{\gamma _{m}}-p_{\gamma _{1}}^{\prime }\cdots
p_{\gamma _{m}}^{\prime }\right) ,
\end{align}%
The collision integrals $\mathcal{A}_{rn}^{\left( \ell \right) }$ can always
be expressed as linear combinations of contractions/projections of $X_{\mu
\nu \gamma _{1}\cdots \gamma _{m}}^{n}$. For the purpose of this paper, we
shall only need $X_{\mu \nu \gamma _{1}\cdots \gamma _{m}}^{n}$ for $m=2$
and $3$. For now we concentrate on calculating these integrals. We separate $%
X_{\mu \nu \gamma _{1}\cdots \gamma _{m}}^{n}$ as%
\begin{equation}
X_{\mu \nu \gamma _{1}\cdots \gamma _{m}}^{n}=A_{\mu \nu \gamma _{1}\cdots
\gamma _{m}}^{n}+B_{\mu \nu \gamma _{1}\cdots \gamma _{m}}^{n}\text{ },
\end{equation}%
with%
\begin{align}
A_{\mu \nu \gamma _{1}\cdots \gamma _{m}}^{n}& =\frac{1}{\nu }\int
dKdK^{\prime }dPdP^{\prime }W_{\mathbf{kk}\prime \rightarrow \mathbf{pp}%
\prime }f_{0\mathbf{k}}f_{0\mathbf{k}^{\prime }}E_{\mathbf{k}}^{n}\text{ }%
k_{\mu }k_{\nu }\left( k_{\gamma _{1}}\cdots k_{\gamma _{m}}+k_{\gamma
_{1}}^{\prime }\cdots k_{\gamma _{m}}^{\prime }\right) ,  \notag \\
B_{\mu \nu \gamma _{1}\cdots \gamma _{m}}^{n}& =-\frac{1}{\nu }\int
dKdK^{\prime }dPdP^{\prime }W_{\mathbf{kk}\prime \rightarrow \mathbf{pp}%
\prime }f_{0\mathbf{k}}f_{0\mathbf{k}^{\prime }}E_{\mathbf{k}}^{n}\text{ }%
k_{\mu }k_{\nu }\left( p_{\gamma _{1}}\cdots p_{\gamma _{m}}+p_{\gamma
_{1}}^{\prime }\cdots p_{\gamma _{m}}^{\prime }\right) .
\end{align}%
The $dPdP^{\prime }$ integration in the first tensor, $A_{\mu \nu \gamma
_{1}\cdots \gamma _{m}}^{n}$, can be immediately performed and written in
terms of the total cross section, $\sigma _{T}(s)$, as%
\begin{equation}
A_{\mu \nu \gamma _{1}\cdots \gamma _{m}}^{n}=\int dKdK^{\prime }f_{0\mathbf{%
k}}f_{0\mathbf{k}^{\prime }}E_{\mathbf{k}}^{n}k_{\mu }k_{\nu }\left(
k_{\gamma _{1}}\cdots k_{\gamma _{m}}+k_{\gamma _{1}}^{\prime }\cdots
k_{\gamma _{m}}^{\prime }\right) \frac{s}{2}\sigma _{T}\left( s\right) .
\end{equation}%
The calculation of the second tensor, $B_{\mu \nu \gamma _{1}\ldots \gamma
_{m}}^{n}$, is cumbersome. First, we write it in the general form%
\begin{equation}
B_{\mu \nu \gamma _{1}\cdots \gamma _{m}}^{n}=-\int dKdK^{\prime }f_{0%
\mathbf{k}}f_{0\mathbf{k}^{\prime }}E_{\mathbf{k}}^{n}k_{\mu }k_{\nu }\Theta
_{\gamma _{1}\cdots \gamma _{m}},
\end{equation}%
where we introduced the tensor%
\begin{equation}
\Theta _{\gamma _{1}\cdots \gamma _{m}}=\frac{2}{\nu }\int dPdP^{\prime }W_{%
\mathbf{kk}\prime \rightarrow \mathbf{pp}\prime }p_{\gamma _{1}}\cdots
p_{\gamma _{m}}.
\end{equation}%
The integral $\Theta _{\gamma _{1}\cdots \gamma _{m}}$ is an $m$-th rank
tensor. Strictly speaking, for isotropic cross sections, this tensor can
only depend on the normalized total momentum of the collision $\tilde{P}%
_{T}^{\mu }\equiv s^{-1/2}\left( k^{\mu }+k^{\prime \mu }\right) \equiv
s^{-1/2}P_{T}^{\mu }$. Thus, the tensor structure of $\Theta _{\gamma
_{1}\cdots \gamma _{m}}$ must be constructed by combinations of $\tilde{P}%
_{T}^{\mu }$ and the projection operator orthogonal to $\tilde{P}_{T}^{\mu }$%
, $\Delta _{P}^{\mu \nu }=g^{\mu \nu }-\tilde{P}_{T}^{\mu }\tilde{P}%
_{T}^{\nu }$. In general,%
\begin{equation}
\Theta _{\gamma _{1}\cdots \gamma _{m}}=\sum_{q=0}^{\left[ m/2\right]
}\left( -1\right) ^{q}a_{mq}\mathcal{C}_{mq}C_{\gamma _{1}\cdots \gamma
_{m}}^{q},
\end{equation}%
where we defined%
\begin{align}
a_{mq}& =\frac{m!}{\left( m-2q\right) !2q!}\left( 2q-1\right) !!,  \notag \\
C_{\gamma _{1}\cdots \gamma _{m}}^{q}& =\Delta _{P}^{(\gamma _{1}\gamma
_{2}}\cdots \Delta _{P}^{\gamma _{2q-1}\gamma _{2q}}\tilde{P}_{T}^{\gamma
_{2q+1}}\cdots \tilde{P}_{T}^{\gamma _{m})},  \notag \\
\mathcal{C}_{mq}& =\frac{2}{\nu \left( 2q+1\right) !!}\int dPdP^{\prime }W_{%
\mathbf{kk}^{\prime }\rightarrow \mathbf{pp}^{\prime }}\left( \tilde{P}%
_{T}^{\mu }p_{\mu }\right) ^{m-2q}\left( -\Delta _{P}^{\alpha \beta
}p_{\alpha }p_{\beta }\right) ^{q}.
\end{align}%
The parentheses $()$ denote the symmetrization of the tensor. For example,%
\begin{align}
\Theta _{\gamma _{1}\gamma _{2}}& =\mathcal{C}_{20}\tilde{P}_{T\gamma _{1}}%
\tilde{P}_{T\gamma _{2}}-\mathcal{C}_{21}\Delta _{P\gamma _{1}\gamma _{2}}, 
\notag \\
\Theta _{\gamma _{1}\gamma _{2}\gamma _{3}}& =\mathcal{C}_{30}\tilde{P}%
_{T\gamma _{1}}\tilde{P}_{T\gamma _{2}}\tilde{P}_{T\gamma _{3}}-\mathcal{C}%
_{31}\left( \Delta _{P\gamma _{1}\gamma _{2}}\tilde{P}_{T\gamma _{3}}+\Delta
_{P\gamma _{1}\gamma _{3}}\tilde{P}_{T\gamma _{2}}+\Delta _{P\gamma
_{2}\gamma _{3}}\tilde{P}_{T\gamma _{1}}\right) \;,
\end{align}%
The integrals $\mathcal{C}_{nq}$ are scalars and can be computed in any
frame. It is most convenient to calculate them in the center-of-momentum
frame, where, $\tilde{P}_{T}^{\mu }=\left( 1,0,0,0\right) $ and $\Delta
_{P}^{\mu \nu }=\mathrm{diag}\left( 0,-1,-1,-1\right) $. Then, it is
straightforward to prove that%
\begin{equation}
\mathcal{C}_{nq}=\frac{\sigma _{T}\left( s\right) }{2^{n}\left( 2q+1\right)
!!}s^{\left( n-2q+1\right) /2}\left( s-4m^{2}\right) ^{\left( 2q+1\right) /2}%
\underset{m\rightarrow 0}{=}\frac{\sigma _{T}\left( s\right) }{2^{n}\left(
2q+1\right) !!}s^{\left( n+2\right) /2}.  \label{Cnq}
\end{equation}%
In the massless limit, the tensors $X_{\mu \nu \gamma _{1}\gamma _{2}}^{n}$
and $X_{\mu \nu \gamma _{1}\gamma _{2}\gamma _{3}}^{n}$ become%
\begin{align}
X_{\mu \nu \gamma _{1}\gamma _{2}}^{n}& =\int dKdK^{\prime }f_{0\mathbf{k}%
}f_{0\mathbf{k}^{\prime }}E_{\mathbf{k}}^{n}k_{\mu }k_{\nu }\sigma
_{T}\left( s\right) k^{\lambda }k_{\lambda }^{\prime }\left( k_{\gamma
_{1}}k_{\gamma _{2}}+k_{\gamma _{1}}^{\prime }k_{\gamma _{2}}^{\prime }-%
\frac{2}{3}P_{T\gamma _{1}}P_{T\gamma _{2}}+\frac{1}{6}sg_{\gamma _{1}\gamma
_{2}}\right) ,  \notag \\
X_{\mu \nu \gamma _{1}\gamma _{2}\gamma _{3}}^{n}& =\int dKdK^{\prime }f_{0%
\mathbf{k}}f_{0\mathbf{k}^{\prime }}E_{\mathbf{k}}^{n}k_{\mu }k_{\nu }\sigma
_{T}\left( s\right) k^{\lambda }k_{\lambda }^{\prime }\left[ k_{\gamma
_{1}}k_{\gamma _{2}}k_{\gamma _{3}}+k_{\gamma _{1}}^{\prime }k_{\gamma
_{2}}^{\prime }k_{\gamma _{3}}^{\prime }-\frac{1}{2}P_{T\gamma
_{1}}P_{T\gamma _{2}}P_{T\gamma _{3}}\right.   \notag \\
& \left. +\frac{1}{6}k^{\beta }k_{\beta }^{\prime }\left( g_{\gamma
_{1}\gamma _{2}}P_{T\gamma _{3}}+g_{\gamma _{1}\gamma _{3}}P_{T\gamma
_{2}}+g_{\gamma _{2}\gamma _{3}}P_{T\gamma _{1}}\right) \right] ,
\end{align}%
where we used that, in the massless limit, $s=2k^{\lambda }k_{\lambda
}^{\prime }$.

\subsection{Particle-diffusion current}

For the collision integrals related to the particle-number diffusion
current, we need the following two contractions 
\begin{align}
\Delta ^{\mu \gamma _{1}}u^{\nu }u^{\gamma _{2}}X_{\mu \nu \gamma _{1}\gamma
_{2}}^{n}& =-\sigma _{T}\left(
I_{10}I_{n+5,1}-4I_{21}I_{n+4,1}-I_{31}I_{n+3,1}\right) ,  \notag \\
\Delta ^{\mu \gamma _{1}}u^{\nu }u^{\gamma _{2}}u^{\gamma _{3}}X_{\mu \nu
\gamma _{1}\gamma _{2}\gamma _{3}}^{n}& =-\frac{\sigma _{T}}{2}\left(
3I_{10}I_{n+6,1}-11I_{21}I_{n+5,1}-5I_{31}I_{n+4,1}-3I_{41}I_{n+3,1}\right) .
\end{align}%
To obtain the above relations, we used Eq.\ (\ref{orthogonality1}) and the
definitions (\ref{Jnq}). In the massless and classical limits the integrals $%
I_{nq}=J_{nq}$ can be calculated analytically%
\begin{equation}
I_{nq}=g\frac{e^{\alpha _{0}}}{\left( 2q+1\right) !!}\frac{1}{2\pi ^{2}}%
\frac{\left( n+1\right) !}{\beta _{0}^{n+2}}=\frac{\left( n+1\right) !}{%
\left( 2q+1\right) !!}\frac{P_{0}}{2\beta _{0}^{n-2}}.  \label{Inq_massless}
\end{equation}%
Then,%
\begin{align}
\Delta ^{\mu \gamma _{1}}u^{\nu }u^{\gamma _{2}}X_{\mu \nu \gamma _{1}\gamma
_{2}}^{-2}& =\frac{4}{3}n_{0}\sigma _{T}\frac{P_{0}}{\beta _{0}},  \notag \\
\Delta ^{\mu \gamma _{1}}u^{\nu }u^{\gamma _{2}}X_{\mu \nu \gamma _{1}\gamma
_{2}}^{0}& =-24n_{0}\sigma _{T}\frac{P_{0}}{\beta _{0}^{3}},  \notag \\
\Delta ^{\mu \gamma _{1}}u^{\nu }u^{\gamma _{2}}u^{\gamma _{3}}X_{\mu \nu
\gamma _{1}\gamma _{2}\gamma _{3}}^{-2}& =12n_{0}\sigma _{T}\frac{P_{0}}{%
\beta _{0}^{2}},  \notag \\
\Delta ^{\mu \gamma _{1}}u^{\nu }u^{\gamma _{2}}u^{\gamma _{3}}X_{\mu \nu
\gamma _{1}\gamma _{2}\gamma _{3}}^{0}& =-280n_{0}\sigma _{T}\frac{P_{0}}{%
\beta _{0}^{4}}.  \label{X1}
\end{align}%
As a consistency check, we confirmed that $\Delta ^{\mu \gamma _{1}}u^{\nu
}u^{\gamma _{2}}X_{\mu \nu \gamma _{1}\gamma _{2}}^{-1}=$ $\Delta ^{\mu
\gamma _{1}}u^{\nu }u^{\gamma _{2}}u^{\gamma _{3}}X_{\mu \nu \gamma
_{1}\gamma _{2}\gamma _{3}}^{-1}=0$.

The components of $\mathcal{A}^{\left( 1\right) }$ change according to the
number of moments included. In the 14-moment approximation, using Eqs.\ (\ref%
{Hk}) and (\ref{Poly}), we obtain%
\begin{equation}
\mathcal{A}_{00}^{\left( 1\right) }=\frac{W^{\left( 1\right) }}{3}%
a_{10}^{(1)}a_{11}^{(1)}\Delta ^{\mu \gamma _{1}}u^{\nu }u^{\gamma
_{2}}X_{\mu \nu \gamma _{1}\gamma _{2}}^{-2}=\frac{4}{9}n_{0}\sigma _{T}.
\end{equation}%
In the 23-moment approximation, e.g.\ considering three polynomials in the
expansion (\ref{expansion2}), for $\ell =1$,%
\begin{align}
\mathcal{A}_{r0}^{\left( 1\right) }& =\frac{W^{\left( 1\right) }}{3}\left[
\left( a_{10}^{(1)}a_{11}^{(1)}+a_{20}^{(1)}a_{21}^{(1)}\right) \Delta ^{\mu
\gamma _{1}}u^{\nu }u^{\gamma _{2}}X_{\mu \nu \gamma _{1}\gamma
_{2}}^{r-2}+a_{20}^{(1)}a_{22}^{(1)}\Delta ^{\mu \gamma _{1}}u^{\nu
}u^{\gamma _{2}}u^{\gamma _{3}}X_{\mu \nu \gamma _{1}\gamma _{2}\gamma
_{3}}^{r-2}\right] ,  \notag \\
\mathcal{A}_{r2}^{\left( 1\right) }& =\frac{W^{\left( 1\right) }}{3}\left(
a_{22}^{(1)}a_{21}^{(1)}\Delta ^{\mu \gamma _{1}}u^{\nu }u^{\gamma
_{2}}X_{\mu \nu \gamma _{1}\gamma _{2}}^{r-2}+a_{22}^{(1)}a_{22}^{(1)}\Delta
^{\mu \gamma _{1}}u^{\nu }u^{\gamma _{2}}u^{\gamma _{3}}X_{\mu \nu \gamma
_{1}\gamma _{2}\gamma _{3}}^{r-2}\right) .
\end{align}%
Then, using the results from Appendix \ref{orthogonal polynomials} for the
coefficients $a_{nq}^{(\ell )}$ together with Eqs.\ (\ref{Inq_massless}) and
(\ref{X1}), we obtain%
\begin{align}
\mathcal{A}_{00}^{\left( 1\right) }& =\frac{2}{3}n_{0}\sigma _{T},\text{ \ \
\ }\mathcal{A}_{02}^{\left( 1\right) }=\frac{\beta _{0}^{2}}{90}n_{0}\sigma
_{T},  \notag \\
\mathcal{A}_{20}^{\left( 1\right) }& =-\frac{4}{3\beta _{0}^{2}}n_{0}\sigma
_{T},\text{ \ \ \ \ }\mathcal{A}_{22}^{\left( 1\right) }=\frac{1}{3}%
n_{0}\sigma _{T}.
\end{align}

\subsection{Shear-stress tensor}

For the collision integrals related to the shear-stress tensor, we need the
following two contractions%
\begin{align}
\Delta ^{\mu \nu \gamma _{1}\gamma _{2}}X_{\mu \nu \gamma _{1}\gamma
_{2}}^{n}& =\frac{10}{3}\sigma _{T}\left(
I_{10}I_{n+5,2}+4I_{21}I_{n+4,2}\right) ,  \notag \\
\Delta ^{\mu \nu \gamma _{1}\gamma _{2}}u^{\gamma _{3}}X_{\mu \nu \gamma
_{1}\gamma _{2}\gamma _{3}}^{n}& =5\sigma _{T}\left(
I_{10}I_{n+6,2}-I_{21}I_{n+5,2}+2I_{31}I_{n+4,2}\right) .
\end{align}%
In order to obtain the above relations, we used Eq.\ (\ref{orthogonality1})
and the definitions (\ref{Jnq}). Using Eq.\ (\ref{Inq_massless}), 
\begin{align}
\Delta ^{\mu \nu \alpha \beta }X_{\mu \nu \alpha \beta }^{-1}& =24\sigma _{T}%
\frac{P_{0}^{2}}{\beta _{0}},  \notag \\
\Delta ^{\mu \nu \alpha \beta }X_{\mu \nu \alpha \beta }^{0}& =\frac{400}{3}%
\sigma _{T}\frac{P_{0}^{2}}{\beta _{0}^{2}},  \notag \\
\Delta ^{\mu \nu \alpha \beta }u^{\gamma _{1}}X_{\mu \nu \alpha \beta \gamma
_{1}}^{-1}& =132\sigma _{T}\frac{P_{0}^{2}}{\beta _{0}^{2}},  \notag \\
\Delta ^{\mu \nu \alpha \beta }u^{\gamma _{1}}X_{\mu \nu \alpha \beta \gamma
_{1}}^{0}& =880\sigma _{T}\frac{P_{0}^{2}}{\beta _{0}^{3}}.  \label{X2}
\end{align}%
The components of $\mathcal{A}^{\left( 2\right) }$ change according to the
number of moments included. In the 14-moment approximation, using Eqs.\ (\ref%
{Hk}) and (\ref{Poly}), we obtain%
\begin{equation}
\mathcal{A}_{00}^{\left( 2\right) }=\frac{W^{\left( 2\right) }}{10}\Delta
^{\mu \nu \gamma _{1}\gamma _{2}}X_{\mu \nu \gamma _{1}\gamma _{2}}^{-1}=%
\frac{3}{5}n_{0}\sigma _{T},
\end{equation}%
where we used Eqs.\ (\ref{Inq_massless}) and (\ref{X2}), together with the
results from Appendix \ref{orthogonal polynomials}.

In the 23-moment approximation, e.g.\ considering two polynomials in the
expansion (\ref{expansion2}), for $\ell =2$,%
\begin{align}
\mathcal{A}_{r0}^{\left( 2\right) }& =\frac{W^{\left( 2\right) }}{10}\left(
1+a_{10}^{(2)}a_{10}^{(2)}\right) \Delta ^{\mu \nu \gamma _{1}\gamma
_{2}}X_{\mu \nu \gamma _{1}\gamma _{2}}^{r-1}+\frac{W^{\left( 2\right) }}{10}%
a_{10}^{(2)}a_{11}^{(2)}\Delta ^{\mu \nu \gamma _{1}\gamma _{2}}u^{\gamma
_{3}}X_{\mu \nu \gamma _{1}\gamma _{2}\gamma _{3}}^{r-1},  \notag \\
\mathcal{A}_{r1}^{\left( 2\right) }& =\frac{W^{\left( 2\right) }}{10}%
a_{11}^{(2)}a_{10}^{(2)}\Delta ^{\mu \nu \gamma _{1}\gamma _{2}}X_{\mu \nu
\gamma _{1}\gamma _{2}}^{r-1}+\frac{W^{\left( 2\right) }}{10}%
a_{11}^{(2)}a_{11}^{(2)}\Delta ^{\mu \nu \gamma _{1}\gamma _{2}}u^{\gamma
_{3}}X_{\mu \nu \gamma _{1}\gamma _{2}\gamma _{3}}^{r-1}.
\end{align}%
Then, using once more the results from Appendix \ref{orthogonal polynomials}
and Eqs.\ (\ref{Inq_massless}) and (\ref{X2}), we obtain%
\begin{align}
\mathcal{A}_{00}^{\left( 2\right) }& =\frac{9}{10}n_{0}\sigma _{T},\text{ \
\ \ }\mathcal{A}_{01}^{\left( 2\right) }=-\frac{1}{20}\beta _{0}n_{0}\sigma
_{T},  \notag \\
\mathcal{A}_{10}^{\left( 2\right) }& =\frac{4}{3\beta _{0}}n_{0}\sigma _{T},%
\text{ \ \ \ }\mathcal{A}_{11}^{\left( 2\right) }=\frac{1}{3}n_{0}\sigma
_{T}.
\end{align}%
We did not calculate the coefficients related to the bulk viscous pressure,
since this quantity vanishes in the massless limit. Also, if the mass was
taken to be finite, some of the steps taken in this appendix would not be
possible.

\section{Transport coefficients}

\label{transport coefficients}

In this appendix we list all the transport coefficients of fluid dynamics
calculated in this paper. The transport coefficients for the bulk viscous
pressure are

\begin{align}
\ell _{\Pi n}& =-\frac{m^{2}}{3}\left( \gamma _{1}^{(1)}\tau _{00}^{\left(
0\right) }-\sum_{r=0,\neq 1,2}^{N_{0}}\tau _{0r}^{\left( 0\right) }\frac{%
G_{3r}}{D_{20}}+\sum_{r=0}^{N_{0}-3}\tau _{0,r+3}^{\left( 0\right) }\Omega
_{r+2,0}^{\left( 1\right) }\right) , \\
\tau _{\Pi n}& =\frac{m^{2}}{3\left( \varepsilon _{0}+P_{0}\right) }\left[
\tau _{00}^{\left( 0\right) }\frac{\partial \gamma _{1}^{(1)}}{\partial \ln
\beta _{0}}-\sum_{r=0,\neq 1,2}^{N_{0}}\tau _{0r}^{\left( 0\right) }\frac{%
G_{3r}}{D_{20}}+\sum_{r=0}^{N_{0}-3}\tau _{0,r+3}^{\left( 0\right) }\beta
_{0}\frac{\partial \Omega _{r+2,0}^{\left( 1\right) }}{\partial \beta _{0}}%
+\sum_{r=0}^{N_{0}-3}\left( r+3\right) \tau _{0,r+3}^{\left( 0\right)
}\Omega _{r+2,0}^{\left( 1\right) }\right] , \\
\delta _{\Pi \Pi }& =\frac{2}{3}\tau _{00}^{\left( 0\right) }+\frac{m^{2}}{3}%
\gamma _{2}^{(0)}\tau _{00}^{\left( 0\right) }\,-\frac{m^{2}}{3}%
\sum_{r=0,\neq 1,2}^{N_{0}}\tau _{0r}^{\left( 0\right) }\frac{G_{2r}}{D_{20}}%
+\frac{1}{3}\sum_{r=0}^{N_{0}-3}\left( r+5\right) \tau _{0,r+3}^{\left(
0\right) }\Omega _{r+3,0}^{\left( 0\right) }-\frac{m^{2}}{3}%
\sum_{r=0}^{N_{0}-5}\left( r+4\right) \tau _{0,r+5}^{\left( 0\right) }\Omega
_{r+3,0}^{\left( 0\right) }  \notag \\
& +\frac{\left( \varepsilon _{0}+P_{0}\right) J_{10}-n_{0}J_{20}}{D_{20}}%
\sum_{r=3}^{N_{0}}\tau _{0r}^{\left( 0\right) }\frac{\partial \Omega
_{r0}^{\left( 0\right) }}{\partial \alpha _{0}}+\frac{\left( \varepsilon
_{0}+P_{0}\right) J_{20}-n_{0}J_{30}}{D_{20}}\sum_{r=3}^{N_{0}}\tau
_{0r}^{\left( 0\right) }\frac{\partial \Omega _{r0}^{\left( 0\right) }}{%
\partial \beta _{0}}, \\
\lambda _{\Pi n}& =-\frac{m^{2}}{3}\left( \tau _{00}^{\left( 0\right) }\frac{%
\partial \gamma _{1}^{(1)}}{\partial \alpha _{0}}+\tau _{00}^{\left(
0\right) }\frac{1}{h_{0}}\frac{\partial \gamma _{1}^{(1)}}{\partial \beta
_{0}}+\sum_{r=0}^{N_{0}-3}\tau _{0,r+3}^{\left( 0\right) }\frac{1}{h_{0}}%
\frac{\partial \Omega _{r+2,0}^{\left( 1\right) }}{\partial \beta _{0}}%
+\sum_{r=0}^{N_{0}-3}\tau _{0,r+3}^{\left( 0\right) }\frac{\partial \Omega
_{r+2,0}^{\left( 1\right) }}{\partial \alpha _{0}}\right) , \\
\lambda _{\Pi \pi }& =-\frac{m^{2}}{3}\left[ -\gamma _{2}^{(2)}\tau
_{00}^{\left( 0\right) }+\sum_{r=0,\neq 1,2}^{N_{0}}\tau _{0r}^{\left(
0\right) }\frac{G_{2r}}{D_{20}}+\sum_{r=0}^{N_{0}-3}\left( r+2\right) \tau
_{0,r+3}^{\left( 0\right) }\Omega _{r+1,0}^{\left( 2\right) }\right] ,
\end{align}%
where $h_{0}=(\varepsilon _{0}+P_{0})/n_{0}$ is the enthalpy per particle.
The transport coefficients for the particle-diffusion current are%
\begin{align}
\delta _{nn}& =\tau _{00}^{\left( 1\right) }+\frac{1}{3}m^{2}\gamma
_{2}^{(1)}\tau _{00}^{\left( 1\right) }-\frac{1}{3}m^{2}\sum_{r=0}^{N_{1}-2}%
\left( r+1\right) \tau _{0,r+2}^{\left( 1\right) }\Omega _{r0}^{\left(
1\right) }+\frac{1}{3}\sum_{r=2}^{N_{1}}\left( r+3\right) \tau _{0r}^{\left(
1\right) }\Omega _{r0}^{\left( 1\right) }  \notag \\
& -\sum_{r=2}^{N_{1}}\tau _{0r}^{\left( 1\right) }\left[ \frac{n_{0}}{D_{20}}%
\left( J_{20}\frac{\partial \Omega _{r0}^{\left( 1\right) }}{\partial \beta
_{0}}+J_{30}\frac{\partial \Omega _{r0}^{\left( 1\right) }}{\partial \alpha
_{0}}\right) -\frac{\varepsilon _{0}+P_{0}}{D_{20}}\left( J_{10}\frac{%
\partial \Omega _{r0}^{\left( 1\right) }}{\partial \beta _{0}}+J_{20}\frac{%
\partial \Omega _{r0}^{\left( 1\right) }}{\partial \alpha _{0}}\right) %
\right] , \\
\ell _{n\Pi }& =\frac{1}{h_{0}}\tau _{00}^{\left( 1\right) }-\gamma
_{1}^{(0)}\tau _{00}^{\left( 1\right) }+\sum_{r=0}^{N_{1}-2}\tau
_{0,r+2}^{\left( 1\right) }\frac{\beta _{0}J_{r+4,1}}{\varepsilon _{0}+P_{0}}%
+\frac{1}{m^{2}}\sum_{r=0}^{N_{1}-2}\tau _{0,r+2}^{\left( 1\right) }\Omega
_{r+3,0}^{\left( 0\right) }-\sum_{r=0}^{N_{1}-4}\tau _{0,r+4}^{\left(
1\right) }\Omega _{r+3,0}^{\left( 0\right) },
\end{align}%
\begin{align}
\tau _{n\Pi }& =\frac{1}{\varepsilon _{0}+P_{0}}\left[ \frac{1}{h_{0}}\tau
_{00}^{\left( 1\right) }-\tau _{00}^{\left( 1\right) }\frac{\partial \gamma
_{1}^{(0)}}{\partial \ln \beta _{0}}+\sum_{r=0}^{N_{1}-2}\tau
_{0,r+2}^{\left( 1\right) }\frac{\beta _{0}J_{r+4,1}}{\varepsilon _{0}+P_{0}}%
+\frac{1}{m^{2}}\sum_{r=0}^{N_{1}-2}\left( r+5\right) \tau _{0,r+2}^{\left(
1\right) }\Omega _{r+3,0}^{\left( 0\right) }\right.  \notag \\
& \left. +\frac{1}{m^{2}}\sum_{r=0}^{N_{1}-2}\tau _{0,r+2}^{\left( 1\right) }%
\frac{\partial \Omega _{r+3,0}^{\left( 0\right) }}{\partial \ln \beta _{0}}%
-\sum_{r=0}^{N_{1}-4}\left( r+4\right) \tau _{0,r+4}^{\left( 1\right)
}\Omega _{r+3,0}^{\left( 0\right) }-\sum_{r=0}^{N_{1}-4}\tau
_{0,r+4}^{\left( 1\right) }\frac{\partial \Omega _{r+3,0}^{\left( 0\right) }%
}{\partial \ln \beta _{0}}\right] , \\
\ell _{n\pi }& =-\gamma _{1}^{(2)}\tau _{00}^{\left( 1\right) }+\frac{1}{%
h_{0}}\tau _{00}^{\left( 1\right) }+\sum_{r=0}^{N_{1}-2}\tau
_{0,r+2}^{\left( 1\right) }\frac{\beta _{0}J_{r+4,1}}{\varepsilon _{0}+P_{0}}%
-\sum_{r=0}^{N_{1}-2}\tau _{0,r+2}^{\left( 1\right) }\Omega _{r+1,0}^{\left(
2\right) }, \\
\tau _{n\pi }& =\frac{1}{\varepsilon _{0}+P_{0}}\left[ \frac{1}{h_{0}}\tau
_{00}^{\left( 1\right) }-\tau _{00}^{\left( 1\right) }\frac{\partial \gamma
_{1}^{(2)}}{\partial \ln \beta _{0}}+\sum_{r=0}^{N_{1}-2}\tau
_{0,r+2}^{\left( 1\right) }\frac{\beta _{0}J_{r+4,1}}{\varepsilon _{0}+P_{0}}%
-\sum_{r=0}^{N_{1}-2}\tau _{0,r+2}^{\left( 1\right) }\frac{\partial \Omega
_{r+1,0}^{\left( 2\right) }}{\partial \ln \beta _{0}}\right.  \notag \\
& \left. -\sum_{r=0}^{N_{1}-2}\left( r+2\right) \tau _{0,r+2}^{\left(
1\right) }\Omega _{r+1,0}^{\left( 2\right) }\right] , \\
\lambda _{nn}& =\frac{3}{5}\tau _{00}^{\left( 1\right) }+\frac{2}{5}%
m^{2}\gamma _{2}^{(1)}\tau _{00}^{\left( 1\right) }-\frac{2}{5}%
m^{2}\sum_{r=0,r\neq 1}^{N_{1}-2}\left( r+1\right) \tau _{0,r+2}^{\left(
1\right) }\Omega _{r0}^{\left( 1\right) }+\frac{1}{5}\sum_{r=2}^{N_{1}}%
\left( 2r+3\right) \tau _{0r}^{\left( 1\right) }\Omega _{r0}^{\left(
1\right) }, \\
\lambda _{n\Pi }& =\tau _{00}^{\left( 1\right) }\left( \frac{1}{h_{0}}\frac{%
\partial \gamma _{1}^{(0)}}{\partial \beta _{0}}+\frac{\partial \gamma
_{1}^{(0)}}{\partial \alpha _{0}}\right) -\frac{1}{m^{2}}%
\sum_{r=0}^{N_{1}-2}\tau _{0,r+2}^{\left( 1\right) }\left( \frac{1}{h_{0}}%
\frac{\partial \Omega _{r+3,0}^{\left( 0\right) }}{\partial \beta _{0}}+%
\frac{\partial \Omega _{r+3,0}^{\left( 0\right) }}{\partial \alpha _{0}}%
\right)  \notag \\
& +\sum_{r=0}^{N_{1}-4}\tau _{0,r+4}^{\left( 1\right) }\left( \frac{1}{h_{0}}%
\frac{\partial \Omega _{r+3,0}^{\left( 0\right) }}{\partial \beta _{0}}+%
\frac{\partial \Omega _{r+3,0}^{\left( 0\right) }}{\partial \alpha _{0}}%
\right) , \\
\lambda _{n\pi }& =\left( \frac{1}{h_{0}}\frac{\partial \gamma _{1}^{(2)}}{%
\partial \beta _{0}}+\frac{\partial \gamma _{1}^{(2)}}{\partial \alpha _{0}}%
\right) \tau _{00}^{\left( 1\right) }+\sum_{r=0}^{N_{1}-2}\tau
_{0,r+2}^{\left( 1\right) }\left( \frac{1}{h_{0}}\frac{\partial \Omega
_{r+1,0}^{\left( 2\right) }}{\partial \beta _{0}}+\frac{\partial \Omega
_{r+1,0}^{\left( 2\right) }}{\partial \alpha _{0}}\right) .
\end{align}

The transport coefficients for the shear-stress tensor are%
\begin{align}
\delta _{\pi \pi }& =\frac{1}{3}m^{2}\gamma _{2}^{(2)}\tau _{00}^{\left(
2\right) }+\frac{1}{3}\sum_{r=0}^{N_{2}}\left( r+4\right) \tau _{0r}^{\left(
2\right) }\Omega _{r0}^{\left( 2\right) }-\frac{1}{3}m^{2}%
\sum_{r=0}^{N_{2}-2}\left( r+1\right) \tau _{0,r+2}^{\left( 2\right) }\Omega
_{r0}^{\left( 2\right) }  \notag \\
& +\sum_{r=0}^{N_{2}}\tau _{0r}^{\left( 2\right) }\left[ \frac{\left(
\varepsilon _{0}+P_{0}\right) J_{10}-n_{0}J_{20}}{D_{20}}\frac{\partial
\Omega _{r0}^{\left( 2\right) }}{\partial \beta _{0}}+\frac{\left(
\varepsilon _{0}+P_{0}\right) J_{20}-n_{0}J_{30}}{D_{20}}\frac{\partial
\Omega _{r0}^{\left( 2\right) }}{\partial \alpha _{0}}\right] , \\
\tau _{\pi \pi }& =\frac{2}{7}\sum_{r=0}^{N_{2}}\left( 2r+5\right) \tau
_{0r}^{\left( 2\right) }\Omega _{r0}^{\left( 2\right) }+\frac{4}{7}%
m^{2}\gamma _{2}^{(2)}\tau _{00}^{\left( 2\right) }-\frac{4}{7}%
m^{2}\sum_{r=0}^{N_{2}-2}\left( r+1\right) \tau _{0,r+2}^{\left( 2\right)
}\Omega _{r0}^{\left( 2\right) }, \\
\lambda _{\pi \Pi }& =\frac{6}{5}\tau _{00}^{\left( 2\right) }+\frac{2}{5}%
m^{2}\gamma _{2}^{(0)}\tau _{00}^{\left( 2\right) }+\frac{2}{5m^{2}}%
\sum_{r=0}^{N_{2}-1}\left( r+5\right) \tau _{0,r+1}^{\left( 2\right) }\Omega
_{r+3,0}^{\left( 0\right) }  \notag \\
& +\frac{2}{5}\sum_{r=3}^{N_{2}}\left( 2r+3\right) \tau _{0r}^{\left(
2\right) }\Omega _{r0}^{\left( 0\right) }-\frac{2}{5}m^{2}\sum_{r=0,\neq
1,2}^{N_{2}-2}\left( r+1\right) \tau _{0,r+2}^{\left( 2\right) }\Omega
_{r0}^{\left( 0\right) }, \\
\;\tau _{\pi n}& =\frac{1}{\varepsilon _{0}+P_{0}}\left[ -\frac{2}{5}%
m^{2}\tau _{00}^{\left( 2\right) }\frac{\partial \gamma _{1}^{(1)}}{\partial
\ln \beta _{0}}+\frac{2}{5}\sum_{r=0}^{N_{2}-1}\left( r+6\right) \tau
_{0,r+1}^{\left( 2\right) }\Omega _{r+2,0}^{\left( 1\right) }-\frac{2}{5}%
m^{2}\sum_{r=0,\neq 1}^{N_{2}-1}\left( r+1\right) \tau _{0,r+1}^{\left(
2\right) }\Omega _{r0}^{\left( 1\right) }\right.  \notag \\
& \left. +\frac{2}{5}\sum_{r=0}^{N_{2}-1}\tau _{0,r+1}^{\left( 2\right) }%
\frac{\partial \Omega _{r+2,0}^{\left( 1\right) }}{\partial \ln \beta _{0}}-%
\frac{2}{5}m^{2}\sum_{r=0}^{N_{2}-3}\tau _{0,r+3}^{\left( 2\right) }\frac{%
\partial \Omega _{r+2,0}^{\left( 1\right) }}{\partial \ln \beta _{0}}\right]
, \\
\ell _{\pi n}& =-\frac{2}{5}m^{2}\gamma _{1}^{(1)}\tau _{00}^{\left(
2\right) }+\frac{2}{5}\sum_{r=0}^{N_{2}-1}\tau _{0,r+1}^{\left( 2\right)
}\Omega _{r+2,0}^{\left( 1\right) }-\frac{2}{5}m^{2}\sum_{r=0,\neq
1}^{N_{2}-1}\tau _{0,r+1}^{\left( 2\right) }\Omega _{r0}^{\left( 1\right) },
\end{align}%
\begin{align}
\lambda _{\pi n}& =-\frac{2}{5}m^{2}\tau _{00}^{\left( 2\right) }\left( 
\frac{1}{h_{0}}\frac{\partial \gamma _{1}^{(1)}}{\partial \beta _{0}}+\frac{%
\partial \gamma _{1}^{(1)}}{\partial \alpha _{0}}\right) +\frac{2}{5}%
\sum_{r=0}^{N_{2}-1}\tau _{0,r+1}^{\left( 2\right) }\left( \frac{1}{h_{0}}%
\frac{\partial \Omega _{r+2,0}^{\left( 1\right) }}{\partial \beta _{0}}+%
\frac{\partial \Omega _{r+2,0}^{\left( 1\right) }}{\partial \alpha _{0}}%
\right)  \notag \\
& -\frac{2}{5}m^{2}\sum_{r=0}^{N_{2}-3}\tau _{0,r+3}^{\left( 2\right)
}\left( \frac{1}{h_{0}}\frac{\partial \Omega _{r+2,0}^{\left( 1\right) }}{%
\partial \beta _{0}}+\frac{\partial \Omega _{r+2,0}^{\left( 1\right) }}{%
\partial \alpha _{0}}\right) .
\end{align}

\section{Calculations}

\label{therm}

In this appendix we compute the quantity $\gamma _{1}^{(2)}$ in the
14-moment approximation and the 23-moment approximation. This variable was
defined in the main text and is given by,%
\begin{equation}
\gamma _{1}^{(2)}=\sum_{n=0}^{N_{2}}\mathcal{F}_{rn}^{\left( 2\right)
}\Omega _{n0}^{\left( 2\right) }.
\end{equation}%
The first step is to compute the thermodynamic integral $\mathcal{F}%
_{rn}^{\left( \ell \right) }$,%
\begin{equation}
\mathcal{F}_{rn}^{\left( \ell \right) }=\frac{\ell !}{\left( 2\ell +1\right)
!!}\int dK\text{ }f_{0\mathbf{k}}\tilde{f}_{0\mathbf{k}}E_{\mathbf{k}}^{-r}%
\mathcal{H}_{\mathbf{k}n}^{\left( \ell \right) }\left( \Delta ^{\alpha \beta
}k_{\alpha }k_{\beta }\right) ^{\ell }.
\end{equation}

\subsection{14-moment approximation}

In this case, $N_{1}=1$ and $N_{2}=0$, and%
\begin{equation}
\gamma _{1}^{(2)}=\mathcal{F}_{10}^{\left( 2\right) }.
\end{equation}%
Also, in the 14-moment approximation,%
\begin{equation*}
\mathcal{H}_{\mathbf{k}0}^{\left( 2\right) }\equiv \frac{W^{\left( 2\right) }%
}{2!}a_{00}^{(2)}P_{0}^{\left( 2\right) }=\frac{W^{\left( 2\right) }}{2!}.
\end{equation*}

In the massless/classical limits%
\begin{equation*}
\mathcal{H}_{\mathbf{k}0}^{\left( 2\right) }=\frac{\beta _{0}^{2}}{8P_{0}},
\end{equation*}%
and finally%
\begin{equation}
\gamma _{1}^{(2)}=\frac{\beta _{0}^{2}}{4P_{0}}\frac{1}{5!!}\int dKf_{0%
\mathbf{k}}E_{\mathbf{k}}^{-1}\left( \Delta ^{\alpha \beta }k_{\alpha
}k_{\beta }\right) ^{2}=\frac{\beta _{0}}{5}.
\end{equation}

\subsection{23-moment approximation}

In this case, $N_{1}=2$ and $N_{2}=1$, and%
\begin{equation}
\gamma _{1}^{(2)}=\mathcal{F}_{10}^{\left( 2\right) }+\Omega _{10}^{\left(
2\right) }\mathcal{F}_{11}^{\left( 2\right) }.
\end{equation}%
Also, in the 23-moment approximation,%
\begin{align}
\mathcal{H}_{\mathbf{k}0}^{\left( 2\right) }& =\frac{W^{\left( 2\right) }}{2!%
}\left( 1+a_{10}^{(2)}P_{1}^{\left( 2\right) }\right) =\frac{W^{\left(
2\right) }}{2!}\left[ 1+\left( a_{10}^{(2)}\right)
^{2}+a_{10}^{(2)}a_{11}^{(2)}E_{\mathbf{k}}\right] ,  \notag \\
\mathcal{H}_{\mathbf{k}1}^{\left( 2\right) }& =\frac{W^{\left( 2\right) }}{2!%
}a_{11}^{(2)}P_{1}^{\left( 2\right) }=\frac{W_{\left( 2\right) }}{2!}\left[
a_{10}^{(2)}a_{11}^{(2)}+\left( a_{11}^{(2)}\right) ^{2}E_{\mathbf{k}}\right]
.
\end{align}%
\newline
We know that%
\begin{equation}
W^{\left( 2\right) }=\frac{\beta _{0}^{2}}{4P_{0}},\text{ \ }\left(
a_{11}^{\left( 2\right) }\right) ^{2}=\frac{\beta _{0}^{2}}{6},\text{ \ }%
\frac{a_{10}^{\left( 2\right) }}{a_{11}^{\left( 2\right) }}=-\frac{6}{\beta
_{0}}.
\end{equation}%
Thus,%
\begin{align}
\mathcal{H}_{\mathbf{k}0}^{\left( 2\right) }& =\frac{\beta _{0}^{2}}{8P_{0}}%
\left( 7-\beta _{0}E_{\mathbf{k}}\right) ,  \notag \\
\mathcal{H}_{\mathbf{k}1}^{\left( 2\right) }& =\frac{\beta _{0}^{3}}{8P_{0}}%
\left( -1+\frac{1}{6}\beta _{0}E_{\mathbf{k}}\right) ,
\end{align}%
and%
\begin{align}
\mathcal{F}_{10}^{\left( 2\right) }& =\frac{\beta _{0}^{2}}{4P_{0}}\frac{1}{%
5!!}\int dK\text{ }f_{0\mathbf{k}}E_{\mathbf{k}}^{-1}\left( 7-\beta _{0}E_{%
\mathbf{k}}\right) \left( \Delta ^{\alpha \beta }k_{\alpha }k_{\beta
}\right) ^{2}=\frac{2}{5}\beta _{0},  \notag \\
\mathcal{F}_{11}^{\left( 2\right) }& =\frac{\beta _{0}^{3}}{4P_{0}}\frac{1}{%
5!!}\int dK\text{ }f_{0\mathbf{k}}E_{\mathbf{k}}^{-1}\left( -1+\frac{1}{6}%
\beta _{0}E_{\mathbf{k}}\right) \left( \Delta ^{\alpha \beta }k_{\alpha
}k_{\beta }\right) ^{2}=-\frac{\beta _{0}^{2}}{30}.
\end{align}

Substituting $\Omega ^{\left( 2\right) }$ from Eq.\ (\ref{matrix1}) we obtain%
\begin{equation}
\gamma _{1}^{(2)}=\frac{2}{15}\beta _{0}=0.133\beta _{0}.
\end{equation}

\section{Orthogonal Polynomials}

\label{orthogonal polynomials}

In this appendix, we construct the set of orthogonal polynomials used in the
main text. These will be polynomials in energy, $E_{\mathbf{k}}=u_{\mu
}k^{\mu }$, i.e., orthogonal polynomials generated by the set $1$, $E_{%
\mathbf{k}}$, $E_{\mathbf{k}}^{2},\ldots $. We construct this orthogonal set
using the Gram-Schmidt orthogonalization method. First we introduce%
\begin{equation}
\omega ^{\left( \ell \right) }\equiv \frac{W^{\left( \ell \right) }}{\left(
2\ell +1\right) !!}\left( \Delta ^{\alpha \beta }k_{\alpha }k_{\beta
}\right) ^{\ell }f_{0\mathbf{k}}\tilde{f}_{0\mathbf{k}}\;,
\end{equation}%
where $f_{0\mathbf{k}}$ is the equilibrium distribution function as defined
in the main text. The weight $W^{(\ell )}$ will be determined such that the
orthogonal polynomial $P_{\mathbf{k}n}^{\left( \ell \right) }$ of order $n=0$
and index $\ell $ is normalized,%
\begin{equation}
\int dK\omega ^{\left( \ell \right) }P_{\mathbf{k}0}^{\left( \ell \right)
}P_{\mathbf{k}0}^{\left( \ell \right) }=1,  \label{ortho}
\end{equation}%
Without loss of generality, the polynomials of order $0$ are set to $1$ for
all values of $\ell $, 
\begin{equation}
P_{\mathbf{k}0}^{\left( \ell \right) }\equiv a_{00}^{(\ell )}=1\;.
\label{orthostart}
\end{equation}%
Then the normalization parameter $W^{(\ell )}$ is obtained from Eq.\ (\ref%
{ortho}),%
\begin{equation}
W^{\left( \ell \right) }=(-1)^{\ell }\frac{1}{J_{2\ell ,\ell }}.  \label{Nl}
\end{equation}%
The thermodynamic functions $J_{nq}$ were defined in the main text, see Eq.\
(\ref{Jnq}).

The polynomials are parametrized as 
\begin{equation}
P_{\mathbf{k}n}^{(\ell )}=\sum_{r=0}^{n}a_{nr}^{(\ell )}E_{\mathbf{k}}^{r}\;.
\label{orthosequence}
\end{equation}%
We construct the polynomials in sequence according to the parametrization (%
\ref{orthosequence}) starting from $n=0$, Eq.~(\ref{orthostart}), using the
orthonormality condition (\ref{conditions}). The orthogonality/normalization
condition implies that, for a polynomial of order $i$, $P_{\mathbf{k}%
i}^{\left( \ell \right) }$,%
\begin{equation}
\int dK\omega ^{\left( \ell \right) }P_{\mathbf{k}i}^{\left( \ell \right)
}P_{\mathbf{k}j}^{(\ell )}=\delta _{ij},
\end{equation}%
for all $j\leq i$. Substituting Eq.\ (\ref{orthosequence}), we obtain the
following equation for the coefficients $a_{ij}^{\left( \ell \right) }$,%
\begin{equation}
\sum_{j=0}^{i}\mathcal{D}_{kj}^{\left( \ell i\right) }\frac{a_{ij}^{\left(
\ell \right) }}{a_{ii}^{\left( \ell \right) }}=\frac{J_{2\ell ,\ell }}{%
\left( a_{ii}^{\left( \ell \right) }\right) ^{2}}\delta _{ki},
\label{EqOrth}
\end{equation}%
where $k=0,\ldots ,i$, and we defined the $\left( i+1\right) \times \left(
i+1\right) $ matrix $\mathcal{D}_{kj}^{\left( \ell i\right) }\equiv
J_{k+j+2\ell ,\ell }$. The solution of Eq.~(\ref{EqOrth}) is 
\begin{align}
\left( a_{ii}^{\left( \ell \right) }\right) ^{2}& =\left( \mathcal{D}%
^{-1}\right) _{ii}^{\left( \ell i\right) }J_{2\ell ,\ell },  \notag \\
\frac{a_{ij}^{\left( \ell \right) }}{a_{ii}^{\left( \ell \right) }}& =\frac{%
\left( \mathcal{D}^{-1}\right) _{ji}^{\left( \ell i\right) }}{\left( 
\mathcal{D}^{-1}\right) _{ii}^{\left( \ell i\right) }},  \label{Coeffss}
\end{align}%
where $\left( \mathcal{D}^{-1}\right) ^{\left( \ell i\right) }$ is the
inverse of $\mathcal{D}^{\left( \ell i\right) }$. For example, for any
polynomial of order $1$, the coefficients are%
\begin{align}
\left( a_{11}^{\left( \ell \right) }\right) ^{2}& =\left( \mathcal{D}%
^{-1}\right) _{11}^{\left( \ell 1\right) }J_{2\ell ,\ell }=\frac{\left(
J_{2\ell ,\ell }\right) ^{2}}{J_{2\ell +2,\ell }J_{2\ell ,\ell }-\left(
J_{2\ell +1,\ell }\right) ^{2}},  \notag \\
\frac{a_{10}^{\left( \ell \right) }}{a_{11}^{\left( \ell \right) }}& =\frac{%
\left( \mathcal{D}^{-1}\right) _{01}^{\left( \ell 1\right) }}{\left( 
\mathcal{D}^{-1}\right) _{11}^{\left( \ell 1\right) }}=-\frac{J_{2\ell
+1,\ell }}{J_{2\ell ,\ell }}.
\end{align}

\section{Irreducible tensors}

\label{irreducible_tensors}

In this appendix, we give some practical relations concerning the
irreducible tensors $k^{\left\langle \mu _{1}\right. }k^{\mu _{2}}\cdots
k^{\left. \mu _{\ell }\right\rangle }$ introduced in the main text. The
definition of these tensors is%
\begin{equation}
k^{\left\langle \mu _{1}\right. }k^{\mu _{2}}\cdots k^{\left. \mu _{\ell
}\right\rangle }=\Delta _{\nu _{1}\cdots \nu _{\ell }}^{\mu _{1}\cdots \mu
_{\ell }}k^{\nu _{1}}\cdots k^{\nu _{\ell }}.
\end{equation}%
The projection operator $\Delta _{\nu _{1}\cdots \nu _{\ell }}^{\mu
_{1}\cdots \mu _{\ell }}$ is symmetric and traceless in the indexes $\mu $
and $\nu $%
\begin{align}
\Delta _{\nu _{1}\cdots \nu _{\ell }}^{\mu _{1}\cdots \mu _{\ell }}& =\Delta
_{\left( \nu _{1}\cdots \nu _{\ell }\right) }^{\left( \mu _{1}\cdots \mu
_{\ell }\right) },  \notag \\
g_{\mu _{i}\mu _{j}}\Delta _{\nu _{1}\cdots \nu _{\ell }}^{\mu _{1}\cdots
\mu _{\ell }}& =g^{\nu _{i}\nu _{j}}\Delta _{\nu _{1}\cdots \nu _{\ell
}}^{\mu _{1}\cdots \mu _{\ell }}=0,\text{ \ }\forall \text{ }1\leq i,j\leq
\ell .
\end{align}%
The parentheses $\left( {}\right) $ on the indices denotes symmetrization of
the tensor. These projections are constructed in Ref.~\cite{DeGroot} and can
be obtained from%
\begin{align}
\Delta ^{\mu _{1}\cdots \mu _{\ell }\nu _{1}\cdots \nu _{\ell }}&
=\sum_{k=0}^{\left[ \ell /2\right] }C\left( \ell ,k\right) \Phi _{(\ell
k)}^{\mu _{1}\cdots \mu _{\ell }\nu _{1}\cdots \nu _{\ell }},  \notag \\
C\left( \ell ,k\right) & =\left( -1\right) ^{k}\frac{\left( \ell !\right)
^{2}}{\left( 2\ell \right) !}\frac{\left( 2\ell -2k\right) !}{k!\left( \ell
-k\right) !\left( \ell -2k\right) !},
\end{align}%
where in the last summation the symbol $\left[ \ell /2\right] $ denotes the
largest integer not exceeding $\ell /2$ and%
\begin{equation}
\Phi _{(\ell k)}^{\mu _{1}\cdots \mu _{\ell }\nu _{1}\cdots \nu _{\ell
}}=\left( \ell -2k\right) !\left( \frac{2^{k}k!}{\ell !}\right)
^{2}\sum_{\wp _{\mu }\wp _{\nu }}\Delta ^{\mu _{1}\mu _{2}}\cdots \Delta
^{\mu _{2k-1}\mu _{2k}}\Delta ^{\nu _{1}\nu _{2}}\cdots \Delta ^{\nu
_{2k-1}\nu _{2k}}\Delta ^{\mu _{2k+1}\nu _{2k+1}}\cdots \Delta ^{\mu _{\ell
}\nu _{\ell }}.
\end{equation}%
This summation is supposed to run over all \textit{distinct} permutations of 
$\mu $--type and $\nu $--type indices (we do not permute the indices $\mu $
with $\nu $). For $\ell =2$ this recipe gives the usual double symmetric and
traceless projection operator $\Delta _{\alpha \beta }^{\mu \nu }$ commonly
employed in relativistic fluid dynamics. As mentioned in the main text, this
set of tensors are useful because they form an orthogonal basis, see Eq.\ (%
\ref{orthogonality1}).

\end{document}